\documentclass[12pt]{article}
\usepackage{graphicx}
\usepackage{natbib}
\usepackage{amsmath}
\usepackage{amssymb}
\usepackage{color}
\usepackage{textcomp}
\usepackage{subfigure}

\newsavebox{\astrutbox}
\sbox{\astrutbox}{\rule[-5pt]{0pt}{20pt}}

\def\D{\partial}

\def\C{{\mathcal{C}}}
\def\A{{\mathcal{A}}}

\def\O{{\mathcal{O}}}
\def\N{{\mathcal{N}}}
\def\P{{\mathcal{P}}}
\def\Q{{\mathcal{Q}}}

\def\E{{\mathcal{E}}}

\def\L{{\mathcal{L}}}
\def\D{{\mathcal{D}}}

\def\T{{\mathcal{T}}}

\def\RR{{\mathbb{R}}}
\def\CC{{\mathbb{C}}}

\def\bA{{\bf A}}
\def\bC{{\bf C}}
\def\bD{{\bf D}}
\def\bF{{\bf F}}

\def\bB{{\bf B}}

\def\bK{{\bf K}}
\def\bM{{\bf M}}

\def\e{{\mathbf e}}

\def\n{{\mathbf n}}
\def\x{{\bm x}}

\def\u{{\bm u}}

\def\t{{\mathbf t}}

\def\x{{\mathbf x}}
\def\y{{\mathbf y}}
\def\z{{\mathbf z}}

\def\b{{\mathbf b}}
\def\u{{\mathbf u}}
\def\v{{\mathbf v}}

\def\0{{\mathbf 0}}

\def\bomega{\boldsymbol{\omega}}
\def\bnabla{\boldsymbol{\nabla}}

\def\Dpartial#1#2{ {\partial #1 \over \partial #2} }

\def\Dpartialn#1#2#3{ {\partial^{#3} #1 \over \partial #2^{#3}} }

\def\Bmp#1{ \begin{minipage}{#1} }
\def\Emp{ \end{minipage} }
\def\Bmpc#1{ \begin{minipage}[c]{#1} }
\def\Bmpt#1{ \begin{minipage}[t]{#1} }
\def\Bmpb#1{ \begin{minipage}[b]{#1} }

\begin{document}
\title{Linear Stability of Hill's Vortex to Axisymmetric Perturbations}
\author{Bartosz Protas$^1$\thanks{Email address for correspondence: bprotas@mcmaster.ca} \ \ and  \
 Alan Elcrat$^2$\thanks{Deceased on December 20, 2013. Alan Elcrat played a key role in the early stages of this study, but did not live to see the final version of the manuscript.}  \\ \\ 
$^1$Department of Mathematics and Statistics, \\ McMaster University, Hamilton, ON, Canada
\\ \\ 
$^2$Department of Mathematics, \\ Wichita State University, Wichita, KS, USA
}

\date{\today}

\maketitle

\begin{abstract}
  We consider the linear stability of Hill's vortex with respect to
  axisymmetric perturbations. Given that Hill's vortex is a solution
  of a free-boundary problem, this stability analysis is performed by
  applying methods of shape differentiation to the contour dynamics
  formulation of the problem in a 3D axisymmetric geometry. This
  approach allows us to systematically account for the effect of
  boundary deformations on the linearized evolution of the vortex
  under the constraint of constant circulation. The resulting singular
  integro-differential operator defined on the vortex boundary is
  discretized with a highly accurate spectral approach. This operator
  has two unstable and two stable eigenvalues complemented by a
  continuous spectrum of neutrally-stable eigenvalues. By considering
  a family of suitably regularized (smoothed) eigenvalue problems
  solved with a range of numerical resolutions we demonstrate that the
  corresponding eigenfunctions are in fact singular objects in the
  form of infinitely sharp peaks localized at the front and rear
  stagnation points. These findings thus refine the results of the
  classical analysis by \citet{mm78}.
\end{abstract}

\begin{flushleft}
Vortex Flows --- Vortex instability;
Mathematical Foundations --- Computational methods;
\end{flushleft}



\section{Introduction}
\label{sec:intro}

Hill's vortex \citep{hill-1894} is one of the few known analytical
solutions of Euler's equations in three-dimensional (3D) space $\Omega
= \RR^3$. In the cylindrical polar coordinate system
$(x,\sigma,\phi)$, it represents a compact axisymmetric region of
azimuthal vorticity $\bomega = [0,0,\omega_{\phi}]$ moving with a
constant velocity along the coordinate direction $x$.  Hill's vortex
is a particular (limiting) case of the Norbury-Fraenkel family of 3D
axisymmetric vortex rings \citep{fraenkel-1972-JFM,norbury-1973-JFM}.
Given the Stokes streamfunction $\psi = \psi(x,\sigma)$ and the
operator $\L := \bnabla\cdot\left(\frac{1}{\sigma}\, \bnabla\right)$,
where $\bnabla := \left[\Dpartial{}{x}, \Dpartial{}{\sigma}\right]^T$
(``:='' means ``equal to by definition''), these flows satisfy the
following system in the frame of reference moving with the translation
velocity $W$ of the vortex
\begin{subequations}
\label{eq:Euler3D}
\begin{alignat}{2}
\L \psi & = - \sigma \, f(\psi) 
&\quad & \textrm{in} \ \Omega, \label{eq:Euler2Da} \\
\psi & \rightarrow  -\frac{1}{2}W \sigma^2 &  & \textrm{as} \ |\x| := \sqrt{x^2+\sigma^2} \rightarrow \infty,
\label{eq:Euler2Db} 
\end{alignat}
\end{subequations}
where the vorticity function $f(\psi)$ has the form
\begin{equation}
f(\psi) = \left\{
\begin{tabular}{ll}
$\C$, \quad &  $\psi > k$\\
0,   \quad & $\psi \le k$
\end{tabular}
\right.,
\label{eq:f}
\end{equation}
in which $\C \neq 0$ and $k \ge 0$ are constants. System
\eqref{eq:Euler3D}--\eqref{eq:f} therefore describes a compact region
$\D$ with azimuthal vorticity $\omega := \omega_{\phi}$ {varying}
proportionally to the distance $\sigma$ from the flow axis embedded in
a potential flow. The boundary of this region $\partial \D := \left\{
  (x,\sigma,\phi)\; : \; \psi(x,\sigma) = k \right\}$ is a priori
unknown and must be found as a part of the problem solution. System
\eqref{eq:Euler3D}--\eqref{eq:f} thus represents a {\em free-boundary}
problem and, {as will become evident below, this property makes
  the study of the stability of solutions more complicated.} For
Hill's vortex the equilibrium shape of the boundary $\partial\D$ has
the form of a sphere with the flow described in the translating frame
of reference by the streamfunction
\begin{equation}
\psi(x,\sigma) =\left\{ \begin{array} {lcl} \vspace{0.2cm}  \dfrac{\C\sigma^2}{10}(a^2- \sigma^2-x^2),& \mbox{if} &  x^2+\sigma^2 \leq a^2, \\
\dfrac{\C \sigma^2 a^2}{15} \left[ \dfrac{a^3}{(x^2+\sigma^2)^{3/2}}-1\right],&
\mbox{if} &  x^2+\sigma^2 > a^2,  \end{array} \right.
\label{eq:psiHill}
\end{equation}
where $a$ is the radius of the sphere. The components of the velocity
field $\v = [v_x, v_{\sigma}, v_{\phi}]^T$ can then be obtained as
$v_x = \frac{1}{\sigma}\Dpartial{\psi}{\sigma}$, $v_{\sigma} =
-\frac{1}{\sigma}\Dpartial{\psi}{x}$ and $v_{\phi} = 0$. Constant $\C$
in \eqref{eq:f}, the translation velocity $W$ and the vortex radius
$a$ are all linked through the relation \citep{wu-book-2006}
\begin{equation}
W = \frac{2}{15} \C a^2
\label{eq:W}
\end{equation}
from which it follows that, for a fixed radius $a$, Hill's vortices
represent a one-parameter family of solutions. To fix attention,
unless indicated otherwise, hereafter we will set $a=1$, $\C=-1$, so
that $W = - 2 / 15$ (i.e., the vortex is moving to the left).

Due to the presence of a sharp interface separating the vortical and
potential flow regions, inviscid vortices in two and three dimensions
are described by equations of the free-boundary type. In addition to
making the process of finding (relative) equilibrium configurations
more difficult, this also complicates their stability analysis. The
main difficulty is that generic perturbations modify the domain on
which the governing partial differential equations (PDEs) are defined
together with their boundary conditions, an effect which must be taken
into account in the derivation of the linearized evolution equations.

Most earlier approaches to studying stability of inviscid vortices
have relied on methods adapted to specific problems. The stability of
the simplest configurations, namely the Rankine and Kirchhoff
vortices, was first investigated, respectively, by \citet{k80} and
\citet{l93}. Further insights about these problems were provided by
the studies of \citet{ms71,b90,ghs04,mr08}.  The linear stability of
more complex vortex configurations, such as polygonal arrays of
corotating vortices and translating vortex pairs, was investigated by
\citet{d85,d90,d95}, see also \citet{dl91}. A noteworthy feature of
their approach is that they also used a {\em continuous} perturbation
equation independent of a particular discretization employed to obtain
the equilibrium solution. The linear stability of the so-called
V-states \citep{woz84}, corotating vortex patches and infinite
periodic arrays of vortices was investigated in detail by \citet{k87},
see as well \citet{saffman-1992}.  This approach was based on the
representation of the solution in terms of the Schwarz function, and
required discretization and numerical differentiation in order to
obtain the perturbation equation. A discrete form of the perturbation
equation was also used by \citet{efm05} in their investigation of the
linear stability of vortices in a symmetric equilibrium with a
circular cylinder and a free stream at infinity.  Another family of
approaches is based on variational energy arguments going back to
Kelvin. They were initially investigated by \citet{d85,ss80,d88,fm08},
and were more recently pursued by \cite{fw10a,fw12b}. They rely on
global properties of the excess energy vs.~velocity impulse diagrams
and provide partial information about the linear stability properties
without the need to actually perform a full linear stability analysis.

A systematic and general approach to the study of stability of
inviscid vortices was recently developed by the present authors
\citep{ep13}. It is defined entirely in the continuous setting and
relies on the ``shape-differential'' calculus \citep{dz01a} for a
rigorous treatment of boundary deformations and their effect on the
linearized evolution. A starting point of this approach is an
equilibrium configuration (a ``fixed point'') of a contour-dynamics
formulation of the flow evolution \citep{p92}. The boundary of the
vortex region is then perturbed in the normal direction and the
contour-dynamics equations are linearized, via shape differentiation,
around the equilibrium configuration. As a result, a singular
integro-differential equation is obtained for the linearized evolution
of the normal perturbation. It is defined on the vortex boundary and
the associated eigenvalue problem encodes information about stability.
In contrast to most of the earlier approaches mentioned above, this
formulation is general, in the sense that it is not tailored to a
particular vortex configuration and does not involve any
simplifications (such as boundary conditions satisfied only
approximately or numerical differentiation). It also does not restrict
the imposed perturbations to be irrotational. Therefore, the obtained
singular integro-differential equation may be considered a
vortex-dynamics analogue of the Orr-Sommerfeld equation used to study
the stability of viscous parallel shear flows \citep{Drazin2004book}.
It was shown by \citet{ep13} that the classical stability analyses of
\citet{k80} and \citet{l93} can be derived as special cases from the
proposed framework. In situations in which the eigenvalue problem is
not analytically tractable, the integro-differential equation can be
approximated numerically with spectral accuracy \citep{ep13}.

As regards the stability of Hill's vortex, which is the subject of the
present study, Moffatt \& Moore made the following remark in their
1978 paper: {\em ``... it is rather remarkable that its stability
  characteristics have not been investigated in detail''}. In the
light of this comment made more than 35 years ago, it is perhaps even
more remarkable that important aspects of this problem in fact still
remain open.  More precisely, only partial results are available
corresponding to the linearized response of Hill's vortex to a
perturbation applied to its boundary. \citet{mm78} studied the
response to axisymmetric perturbations by analyzing an approximate
equation for the evolution of the vortex boundary (a similar approach
had been developed earlier by \citet{b73}). They key finding was that
the perturbations evolve towards the shape of a sharp ``spike''
localized at the rear stagnation point and directed into or out of the
vortex depending on the form of the initial perturbation. The
authors also noted the absence of any oscillatory components in the
vortex response.  These observations were confirmed by the
computations of \citet{p86} who studied the evolution of disturbances
well into the nonlinear regime.  He demonstrated that the spike-like
deformation of the vortex boundary arising from the linear instability
leads to a significant fluid detrainment or entrainment over longer
times. The response of Hill's vortex to general, non-axisymmetric
perturbations was studied using approximate techniques based on
expansions of the flow variables in spherical harmonics and numerical
integration by \citet{frk94,r99}.  Their main findings were consistent
with those of \citet{mm78}, namely, that perturbations of the vortex
boundary develop into sharp spikes whose number depends on the
azimuthal wavenumber of the perturbation.  A number of investigations
\citep{l95,rf00,hh10} studied the stability of Hill's vortex with
respect to short-wavelength perturbations applied locally and advected
by the flow in the spirit of the WKB approach \citep{lh91}. These
analyses revealed the presence of a number of instability mechanisms,
although they are restricted to the short-wavelength regime. In this
context we also mention the study by \citet{lsf01} who considered the
linear response of the compressible Hill's vortex to acoustic waves.

Our present investigation attempts to complete the picture by
performing, for a first time, a {\em global} spectral stability
analysis of Hill's vortex with respect to axisymmetric perturbations.
We provide numerical evidence based on highly accurate computations
that this problem is not, in fact, well posed in the sense that the
eigenfunctions are ``distributions'' (i.e., they are not continuous
functions of the arclength). By considering a sequence of suitably
regularized problems and using methods of harmonic analysis it is
demonstrated that the eigenfunctions corresponding to, respectively,
the unstable and stable modes have the form of infinitely sharp spikes
localized at the rear and front stagnation points. These findings thus
refine the conclusions from the earlier approximate computations of
\citet{mm78,frk94,r99}. We also show that the discrete spectrum
corresponding to the stable and unstable modes is complemented by a
continuous spectrum of equally non-smooth neutrally-stable eigenmodes.
The structure of the paper is as follows: in the next section we use
methods of the shape calculus to derive the stability equation; in \S
\ref{sec:numer} we describe and validate the numerical approach,
whereas computational results are presented in \S \ref{sec:results};
{the results are discussed in \S \ref{sec:discuss}} and final
comments are deferred to \S \ref{sec:final}.

\section{Derivation of the Stability Equation}
\label{sec:stab}

In this section we first provide details of the contour-dynamics
formulation in the 3D axisymmetric geometry which is the basis of our
approach. Then, methods of shape calculus are used to derive an
integro-differential equation characterizing the stability of Hill's
vortex. Finally, we discuss some properties of this equation.
Hereafter $\A$ will denote the projection of the axisymmetric vortex
region $\D$ onto the meridional plane $\{x,\sigma\}$.

The formalism of contour dynamics is a convenient way to study the
evolution of inviscid flows with piecewise smooth vorticity
distributions \citep{p92}. Given a time-dependent region $\A(t)$,
where $t$ is time, its evolution can be studied by tracking the points
$\y(t)$ on its boundary $\partial\A(t)$ via the equation
\begin{equation}
\frac{d\y(t)}{dt} = \v(\y(t)) = \C \, \int_{\partial\A(t)} \bK(\y(t),\y')\, ds_{\y'}, \quad
\forall \y(t) \in \partial\A(t),
\label{eq:v}
\end{equation}
where $\bK(\y(t),\y')$ is a suitable Biot-Savart kernel, $\y$ and
$\y'$ are defined in the absolute frame of reference and $ds_{\y'}$ is
an arclength element of the vortex boundary in the meridional plane.
An equilibrium shape of the vortex boundary, denoted $\partial\A$
(without the argument $t$), can be characterized by transforming the
coordinates to the translating frame of reference $\x(t) := \y(t) - W
t\, \e_x$ and considering the normal component of equation
\eqref{eq:v}
\begin{equation}
\n_{\x}\cdot\frac{d\x(t)}{dt} = \C \, \n_{\x}\cdot\left[
\int_{\partial\A} \bK(\x(t),\x')\, ds_{\x'} - W\, \e_x\right] = 0, 
\label{eq:vn}
\end{equation}
where $\n_{\x}$ denotes the unit vector normal to the contour
$\partial\A$ at the point $\x$ (hereafter we will use the convention
that a subscript on a geometric quantity will indicate where this
quantity is evaluated). Since we have $\x(0) = \y(0)$, the arguments
of the kernel $\bK$ in \eqref{eq:vn} can be changed to $\x$ and $\x'$.
Equation \eqref{eq:vn} expresses the vanishing of the normal velocity
component on the vortex boundary in relative equilibrium. Since the
equilibrium shape of the boundary $\partial\A$ is in general a priori
unknown, relation \eqref{eq:vn} reveals the free-boundary aspect of
the problem. The Biot-Savart kernel was derived by {\citet{wr96}}
with an alternative, but equivalent, formulation also given by
\citet{p86}. {Here we will use this kernel in the form rederived
  and tested by \citet{slf08}}
\begin{equation}
\begin{aligned}
\bK(x\,\e_x+\sigma\,\e_{\sigma}, x'\,\e_x+\sigma'\,\e_{\sigma}) =
& \left[ (x'-x) \, G(x,x',\sigma,\sigma') \, \cos\theta' - 
\sigma \, H(x,x',\sigma,\sigma') \, \sin\theta'\right]\, \e_x + \\
& \sigma'  H(x,x',\sigma,\sigma') \, \cos\theta' \, \e_{\sigma},
\end{aligned}
\label{eq:K}
\end{equation}
where 
\begin{align*}
G(x,x',\sigma,\sigma') &:=  \frac{\sigma'}{\pi \sqrt{A+B}} K(\tilde{r}), \\
H(x,x',\sigma,\sigma') &:=  \frac{1}{2\pi\sigma} \left[ \frac{A}{\sqrt{A+B}} K(\tilde{r}) - E(\tilde{r})\, \sqrt{A+B}\right]
\end{align*}
in which 
\begin{equation*}
A := (x-x')^2 + \sigma^2 + \sigma'^2, \qquad B := 2 \sigma\sigma', \qquad \tilde{r} := \sqrt{\frac{2B}{A+B}},
\end{equation*}
whereas $\theta'$ denotes the polar angle of the point $\x'$, i.e.,
$\cos\theta' = x' / \sqrt{x'^2 + \sigma'^2}$ and $\sin\theta' =
\sigma' / \sqrt{x'^2 + \sigma'^2}$, and $K(\tilde{r})$ and
$E(\tilde{r})$ are the complete elliptic integrals of the first and
second kind, respectively \citep{olbc10}. We note that $\tilde{r}
\rightarrow 1$ as $x' \rightarrow x$ and $\sigma' \rightarrow \sigma$,
and at $\tilde{r}=1$ {function} $K(\tilde{r})$ has a logarithmic
singularity (more details about the singularity structure of kernel
\eqref{eq:K} will be provided below).

In addition to the circulation, impulse and energy conserved by all
classical solutions of Euler's equations, axisymmetric inviscid flows
also conserve {Casimirs $\iint_{\A(t)} \Phi({\omega} / {\sigma})
  \,\sigma \, d\A$, where $\Phi \; : \; \RR \rightarrow \RR$ is an
  arbitrary function with sufficient regularity
  \citep{mohseni-2001-PF}.}  Since circulation is particularly
important from the physical point of view, we will focus on stability
analysis with respect to perturbations {preserving this quantity, 
which is the same approach as was also taken by \citet{mm78}.}
Circulation $\Gamma$ of the flow in the meridional plane is defined as
\begin{equation}
\Gamma := \iint_{\A(t)} {\omega} \, d\A
\label{eq:circ}
\end{equation}
{and can be also viewed as the Casimir corresponding to
  $\Phi(\xi) = \xi$.}  We add that, since the flows considered here
have the property $\omega = \C \sigma$,
cf.~\eqref{eq:Euler3D}--\eqref{eq:f}, conservation of circulation
\eqref{eq:circ} implies the conservation of the volume of the vortical
region.

Following the ideas laid out by \citet{ep13}, we introduce a
parameterization of the contour $\x = \x(t,s) \in \partial A(t)$ in
terms of its arclength $s$. We will also adopt the convention that the
superscript $\epsilon$, where $0 < \epsilon \ll 1$, will denote
quantities corresponding to the perturbed boundary, so that $\x =
\x^{\epsilon}|_{\epsilon = 0}$ and $\n_{\x} = \n_{\x^{\epsilon}}|_{\epsilon =
  0}$ are the quantities corresponding to the (relative) equilibrium.
Then, points on the perturbed vortex boundary can be represented as
follows
\begin{equation}
\x^{\epsilon}(t,s) = \x(s) + \epsilon\, r(t,s) \, \n_{\x}(s),
\label{eq:x}
\end{equation}
where $r(t,s)$ represents the ``shape'' of the perturbation. We note
that, without affecting the final result, $\n_{\x}(s)$ in the last
term in \eqref{eq:x} could be replaced with its perturbed counterpart
$\n_{\x^{\epsilon}}(t,s)$. Using equation \eqref{eq:v} we thus deduce
\begin{equation}
\n_{\x^{\epsilon}}\cdot\frac{d\x^{\epsilon}(t)}{dt} = 
{\n_{\x^{\epsilon}}\cdot}\v^{\epsilon}(\x^{\epsilon}(t)) = 
\C \, \n_{\x^{\epsilon}}\cdot\left[
\int_{\partial\A^{\epsilon}(t)} \bK(\x^{\epsilon}(t),\x')\, ds_{\x'} - W\, \e_x\right], 
\label{eq:vne}
\end{equation}
from which the equilibrium condition \eqref{eq:vn} is obtained by
setting $\epsilon = 0$. The perturbation equation is obtained by
linearizing relation \eqref{eq:vne} around the equilibrium
configuration characterized by \eqref{eq:vn} which is equivalent to
expanding \eqref{eq:vne} in powers of $\epsilon$ and retaining the
first-order terms. Since equation \eqref{eq:vne} involves perturbed
quantities defined on the perturbed vortex boundary $\partial
A^{\epsilon}(t)$, the proper way to obtain this linearization is using
methods of the shape-differential calculus \citep{dz01a}. Below we
state the main results only and refer the reader to our earlier study
\citep{ep13} for details of all intermediate transformations.
Shape-differentiating the left-hand side (LHS) of relation
\eqref{eq:vne} and setting $\epsilon = 0$ we obtain
\begin{equation}
\frac{d}{d\epsilon} \left[\n_{\x^{\epsilon}}\cdot\frac{d\x^{\epsilon}(t)}{dt}\right]\bigg|_{\epsilon=0} = \Dpartial{r}{t}.
\label{eq:vne_lhs}
\end{equation}
As regards the right-hand side (RHS) in \eqref{eq:vne}, we obtain
\begin{multline}
\frac{d}{d\epsilon} \left\{ \C \, \n_{\x^{\epsilon}}\cdot\left[
\int_{\partial\A^{\epsilon}(t)} \bK(\x^{\epsilon}(t),\x')\, ds_{\x'} - W\, \e_x\right] \right\}\Bigg|_{\epsilon=0} = \\
= - \Dpartial{r}{s} \, \v_0\cdot\t_{\x} 
+\C \, r(s)\, \n_{\x}\cdot\int_{\partial\A} \Dpartial{\bK}{n_{\x}}\, ds' 
+\C \, \n_{\x}\cdot \int_{\partial\A} \left[\Dpartial{\bK}{n_{\x'}} + \varkappa_{\x'}\bK\right] r(s')\, ds',
\label{eq:vne_rhs}
\end{multline}
where $\t_{\x}$ is the unit tangent vector and $\varkappa_{\x}$ the
curvature of the contour $\partial\A$ (in the present case, with the
contour $\partial\A$ given by a half-circle of unit radius, $\varkappa
\equiv 1$). The orientation of the unit vectors $\t_{\x}$ and
$\n_{\x}$, and the sign of the curvature $\varkappa_{\x}$ satisfy
Frenet's convention. As explained by \citet{ep13}, the three terms on
the RHS of \eqref{eq:vne_rhs} represent the shape-sensitivity of the
RHS of \eqref{eq:vne} to perturbations \eqref{eq:x} applied separately
to the normal vector $\n_{\x}$, the evaluation point $\x$ and the
contour $\partial\A$ over which the integral is defined. Since the
flow evolution is subject to constraint \eqref{eq:circ}, this will
restrict the admissible perturbations $r$. Indeed, noting that $\omega
= \C\sigma$, cf.~\eqref{eq:Euler3D}--\eqref{eq:f}, and
shape-differentiating relation \eqref{eq:circ} we obtain the following
condition \citep{ep13}
\begin{equation}
\int_{\partial\A} {\sigma} r(s')\, ds' = 0
\label{eq:r0}
\end{equation}
{restricting the class of admissible perturbations to those which
  do not modify the circulation \eqref{eq:circ}, although the
  vorticity may change locally (analogous conditions {can} be obtained
  when the perturbations are constructed to preserve other integral
  invariants mentioned above).}  In the case of Hill's vortex with the
assumed parameter values, the equilibrium shape of the vortex boundary
is given by the sphere of unit radius, so that $\partial\A = \{
(x,\sigma) \; : \; x^2 + \sigma^2 = 1 \}$. In such case the arclength
coordinate $s$ reduces to the polar angle $\theta \in [0,\pi]$, so
that $x = \cos\theta$ and $\sigma = \sin\theta$. Therefore, below we
will use $\theta$ as our independent variable.

Combining \eqref{eq:vne_lhs}--\eqref{eq:r0} and replacing the line
integrals with the corresponding definite ones we finally obtain the
perturbation equation
\begin{subequations}
\label{eq:L}
\begin{align}
 \Dpartial{r}{t} & = - \Dpartial{r}{\theta} \, \v_0\cdot\t_{\theta} 
+\C \, r(\theta)\, \int_0^{\pi} I_2(\theta,\theta')\, d\theta' + \C \, \int_0^{\pi} I_1(\theta,\theta')r(\theta')\, d\theta' 
\label{eq:La} \\
& := \left( \L r \right)(\theta) \nonumber \\
& \textrm{subject to:} \quad \int_0^{\pi} {\sin\theta'} r(\theta')\, d\theta' = 0,
\label{eq:Lb}
\end{align}
\end{subequations}
where $\L$ denotes the associated linear operator and
\begin{subequations}
\label{eq:I}
\begin{align}
I_1(\theta,\theta') & := \n_\theta \cdot \left[\Dpartial{\bK(\theta,\theta')}{n_{\theta'}} + \varkappa_{\theta'}\bK(\theta,\theta')\right] = 
\left[-3\cos\theta + 5 \,\cos\theta'\right] \, \Q(\theta,\theta'), \label{eq:I1} \\
I_2(\theta,\theta') & := \n_\theta \cdot \Dpartial{\bK(\theta,\theta')}{n_{\theta}} = 
\left[\cos\theta +\cos\theta' \right] \, \Q(\theta,\theta') \label{eq:I2} 
\end{align}
\end{subequations}
in which 
\begin{align*}
\Q(\theta,\theta') &:= {\frac { \left[\cos\theta' \sin(-\theta + \theta') + \sin\theta -\sin\theta' \right] K( \tilde{R}) 
+\sin\theta'  \left[ \cos\theta -\cos\theta' \right]^2 E( \tilde{R}) }
{2 \pi \, \left[ \cos(\theta-\theta') -1 \right] \sqrt {2- 2\,\cos(\theta-\theta')}}}, \\
\tilde{R} &:= \sqrt {{\left[\cos \left( -\theta+\theta' \right) 
-\cos \left( \theta+\theta' \right)\right]
 / \left[1-\cos \left( \theta+\theta' \right) \right]}}.
\end{align*}
As regards the singularities of the kernels, one can verify by
inspection that
\begin{subequations}
\label{eq:Ksing}
\begin{align}
\forall \ \theta \in (0,\pi) \quad
& \lim_{\theta' \rightarrow \theta} \n_\theta \cdot \bK(\theta,\theta') = 0, \label{eq:Ksing_a} \\
& \lim_{\theta' \rightarrow \theta} \frac{\t_\theta \cdot \bK(\theta,\theta')}{\ln|\theta-\theta'|} = 
- \frac{1}{2\pi} \sin \theta, \label{eq:Ksing_b} \\
& \lim_{\theta' \rightarrow \theta} \frac{I_1(\theta,\theta')}{\ln|\theta-\theta'|} = 
- \frac{1}{2\pi} \cos \theta, \label{eq:Ksing_c} \\
& \lim_{\theta' \rightarrow \theta} \frac{I_2(\theta,\theta')}{\ln|\theta-\theta'|} = 
\frac{1}{2\pi} \cos \theta. \label{eq:Ksing_d}
\end{align}
\end{subequations}
The singularities of the kernels $I_1(\theta,\theta')$ and
$I_2(\theta,\theta')$ vanish at $\theta=0$ and $\theta=\pi$.
Properties \eqref{eq:Ksing_a}--\eqref{eq:Ksing_d} will be instrumental
in achieving spectral accuracy in the discretization of system
\eqref{eq:L}.

Equation \eqref{eq:La} is a first-order integro-differential equation
and as such would in principle require only one boundary condition.
However, since the kernel \eqref{eq:K} is obtained via averaging with
respect to the azimuthal angle $\phi$ (due to the axisymmetry
assumption, see \cite{slf08}), the different terms and integrands in
equation \eqref{eq:La} exhibit the following reflection symmetries
\begin{subequations}
\label{eq:refl}
\begin{align}
\forall \ \theta \in [0,\pi] \quad
\left(\v_0\cdot\t\right)_\theta &= - \left(\v_0\cdot\t\right)_{-\theta}, \label{eq:refl_a} \\
I_1(\theta,\theta') &= - I_1(-\theta,-\theta'),  \label{eq:refl_b} \\
I_2(\theta,\theta') &= - I_2(-\theta,-\theta')  \label{eq:refl_c}
\end{align}
\end{subequations}
indicating that equation \eqref{eq:La} is invariant with respect to
the change of the independent variable $\eta = -\theta \in [-\pi,0]$.
This means that when equation \eqref{eq:La} is considered on an
extended periodic domain $[-\pi,\pi]$ subject to {\em even} initial
data $r(0,\theta) = r(0,-\theta)$, $\theta \in (0,\pi)$, its solution
will also remain an even function of $\theta$ at all times, i.e.,
$r(t,\theta) = r(t,-\theta)$, $t>0$. In particular, {\em if}\;\;they
are smooth enough, these solutions will satisfy the symmetry
conditions
\begin{equation}
\Dpartialn{r}{\theta}{2k-1}\bigg|_{\theta=0} = \Dpartialn{r}{\theta}{2k-1}\bigg|_{\theta=\pi} = 0,
\quad k=1,2,\dots.
\label{eq:BCsym}
\end{equation}
Thus, system \eqref{eq:L} with even initial data (which is consistent
with the axisymmetry assumption) and subject to boundary conditions
\eqref{eq:BCsym} is {\em not} an over-determined problem. These
observations will be used in the next section to construct a
spectral discretization of equation \eqref{eq:La}.

After introducing the ansatz 
\begin{equation}
{r(t,\theta) = e^{i \lambda t}\,u(\theta) + C.C.,} 
\label{eq:r}
\end{equation}
where $i:=\sqrt{-1}$ and $\lambda \in \CC$, system \eqref{eq:L}
together with the boundary conditions \eqref{eq:BCsym} takes the form
of a constrained eigenvalue problem
\begin{subequations}
\label{eq:evalp}
\begin{alignat}{2}
& i \, \lambda \, u(\theta) = \left( \L u\right)(\theta), & \quad & \theta \in (0,\pi) \label{eq:evalp1} \\
& \textrm{subject to:} \quad \Dpartialn{u}{\theta}{2k-1}\bigg|_{\theta=0} = \Dpartialn{u}{\theta}{2k-1}\bigg|_{\theta=\pi} = 0, &
\qquad & k=1,2,\dots, \label{eq:evalp2} \\
& \phantom{\textrm{subject to:}} \quad \int_0^{\pi} {\sin\theta'}u(\theta')\, d\theta' = 0, & & \label{eq:evalp3}
\end{alignat}
\end{subequations}
where the operator $\L$ is defined in \eqref{eq:La}. The
eigenvalues $\lambda$ and the eigenfunctions $u$ characterize the
stability of Hill's vortex to axisymmetric perturbations.

\section{Numerical Approach}
\label{sec:numer}

In this section we describe the numerical approach with a focus on the
discretization of system \eqref{eq:evalp} and the solution of the
resulting algebraic eigenvalue problem. We will also provide some
details about how this approach has been validated. We are interested
in achieving the highest possible accuracy and, in principle,
eigenvalue problems for operators defined in the continuous setting on
one-dimensional (1D) domains can be solved with machine precision
using {\tt chebfun} \citep{chebfun}. However, at present, {\tt
  chebfun} does not have the capability to deal with singular integral
operators such as $\L$. We have therefore implemented an alternative
hybrid approach relying on a representation of the operator $\L$ in a
trigonometric basis in which kernel singularities are treated
analytically and {\tt chebfun} is used to evaluate the remaining
definite integrals with high precision.

The eigenfunctions $u$ are approximated with the following truncated
series
\begin{equation}
u(\theta) \approx u^N(\theta) := \sum_{k=0}^{N-1} \alpha_k \cos k\theta,
\label{eq:un}
\end{equation}
where $\alpha_0,\dots,\alpha_{N-1} \in \RR$ are unknown coefficients,
which satisfies exactly the boundary conditions \eqref{eq:BCsym}. The
interval $[0,\pi]$ is discretized with equispaced grid points
\begin{equation}
\theta_j = \frac{\pi}{N-1} j, \quad j=0,\dots,N-1
\label{eq:tj}
\end{equation}
(both endpoints are included in the grid). After substitution of
ansatz \eqref{eq:un}, {equation} \eqref{eq:evalp1} is collocated
on grid \eqref{eq:tj}, {whereas constraint \eqref{eq:evalp3} takes
  the form
\begin{equation}
\sum_{k=0}^{N-1} \alpha_k 
\int_0^{\pi} \sin\theta'\cos k\theta'\, d\theta' = 0
\label{eq:dcirc}
\end{equation}
in which the integrals are evaluated as
\begin{equation}
\int_0^{\pi} \sin\theta'\cos k\theta'\, d\theta' = 
\left\{ 
\begin{alignedat}{2}
& 0, & \quad & k = 1,3,5,\dots \\
- & \frac{2}{k^2 - 1}, && k = 0,2,4,\dots
\end{alignedat}
\right.
\label{eq:sincos}
\end{equation}
(we note that these integrals vanish for all odd values of $k$). As a
result, we obtain the following discrete eigenvalue problem}
\begin{subequations}
\label{eq:evald}
\begin{align}
i \lambda \sum_{k=0}^{N-1} A_{jk}\alpha_k 
& = \sum_{k=0}^{N-1} \left( B_{jk} + C_{jk} + D_{jk} \right) \alpha_k,
\quad  \quad j=0,\dots,N-1, \label{eq:evalda} \\
\mathop{\sum_{k=0}^{N-1}}_{k \; \text{even}} \frac{\alpha_k}{k^2 - 1} & = 0 \label{eq:evaldb}
\end{align}
\end{subequations}
in which 
\begin{equation}
A_{jk} := \cos k\theta_j, \qquad j,k=0,\dots,N-1
\label{eq:Ajk}
\end{equation}
is the (invertible) collocation matrix, whereas the matrices $\bB$,
$\bC$ and $\bD$ correspond to the three terms in operator $\L$,
cf.~\eqref{eq:La}. We remark that the {terms corresponding to all
  values $k=0,\dots,N-1$ have} to be included in expansion
\eqref{eq:un}, even though {their sum} might not satisfy
constraint \eqref{eq:Lb}, as otherwise the collocation problem is not
well posed (i.e., matrix \eqref{eq:Ajk} is singular). Constraint
{\eqref{eq:r0}} is then imposed through the generalized
formulation \eqref{eq:evald}.

The entries of matrix $\bB$, corresponding to the first term in
operator $\L$, are defined as follows
\begin{equation}
B_{jk} := (\v_0\cdot\t)_{\theta_j} \, k \, \sin k\theta_j, \qquad j,k=0,\dots,N-1.
\label{eq:Bjk}
\end{equation}
The entries of matrix $\bC$, corresponding to the second term in
operator $\L$, are defined as follows
\begin{equation}
C_{jk} := \left(\C\, \int_0^{\pi} I_2(\theta_j,\theta')\, d\theta'\right) \, A_{jk}, 
\qquad j,k=0,\dots,N-1,
\label{eq:Cjk}
\end{equation}
where the coefficient is given by an improper integral evaluated using
the property \eqref{eq:Ksing_d} to separate the singular part of the
kernel \citep{h95}
\begin{equation}
\int_0^{\pi} I_2(\theta,\theta')\, d\theta' =: 
\underbrace{\int_0^{\pi} \left[I_2(\theta,\theta') - \frac{\cos\theta}{2\pi} \ln|\theta-\theta'|\right]\, d\theta'}_{\T(\theta)} +  
\underbrace{\frac{\cos\theta}{2\pi} \int_0^{\pi}\ln|\theta-\theta'|\, d\theta'}_{\mathcal{S}(\theta)},
\label{eq:intI2}
\end{equation}
where $\T(\theta)$ has a bounded and continuous integrand expression
and can be therefore accurately evaluated using the function {\tt sum}
from {\tt chebfun} (for $\theta' \approx \theta$ the integrand is
represented in terms of a generalized series expansion), whereas
$\mathcal{S}(\theta)$ can be computed analytically as
\begin{equation}
\mathcal{S}(\theta) = \frac{\cos\theta}{2\pi} \left[
(\pi - \theta)\, \ln(\pi-\theta) - \theta\,\ln\theta - \pi \right], 
\qquad \theta \in [0,\pi].
\label{eq:S}
\end{equation}
The entries of matrix $\bD$, corresponding to the last term in
operator $\L$, are defined as follows
\begin{equation}
D_{jk} := \C\, \int_0^{\pi} I_1(\theta_j,\theta')\cos k\theta'\, d\theta', 
\qquad j,k=0,\dots,N-1
\label{eq:Djk}
\end{equation}
which represents the action of a weakly-singular integral operator on
trigonometric functions. In the light of the property
\eqref{eq:Ksing_c}, they are evaluated similarly to \eqref{eq:intI2}
by separating the singular part of the kernel. We thus obtain
\begin{equation}
\begin{aligned}
\int_0^{\pi} I_1(\theta,\theta')\cos(k\theta')\, d\theta' =: 
&\underbrace{\int_0^{\pi} \ln|\theta-\theta'| 
\left[ \frac{I_1(\theta,\theta')}{\ln|\theta-\theta'|} + \frac{\cos\theta}{2\pi} \right] \cos k\theta'\, d\theta'}_{\T_k(\theta)} \\ 
& - \underbrace{\frac{\cos\theta}{2\pi} \int_0^{\pi}\ln|\theta-\theta'|\cos k\theta'\, d\theta'}_{\mathcal{S}_k(\theta)},
\end{aligned}
\label{eq:intI1}
\end{equation}
where $\T_k(\theta)$ has a bounded and continuous integrand expression
and can be therefore accurately evaluated using the function {\tt sum}
from {\tt chebfun} (for $\theta' \approx \theta$ the integrand is
represented in terms of a generalized series expansion). As regards
$\mathcal{S}_k(\theta)$, for $k=0$, it is already given in
\eqref{eq:S}. For $k>0$, we obtain
\begin{equation}
\begin{aligned}
\int_0^{\pi}\ln|\theta-\theta'|\cos(k\theta')\, d\theta' =
\frac{i}{2k} \big\{ 
& e^{ik\theta} \left[ \E_1(ik\theta) - \E_1(ik(\theta-\pi)) + \pi i\right] \\
- & e^{-ik\theta} \left[ \E_1(-ik\theta) - \E_1(ik(\pi-\theta)) - \pi i\right]\big\},
\end{aligned}
\label{eq:Sk}
\end{equation}
where $\E_1(z)$, $z\in \CC$, is the exponential integral {defined
  as the complex extension of the function $\E_1(x) := \int_x^{\infty}
  ( e^{-t} / t ) \, dt$, $x>0$} \citep{olbc10}. We note that, while
the LHS of relation \eqref{eq:Sk} is real-valued, it is evaluated in
terms of a combination of complex-valued expressions. Since the
exponential integral is multi-valued in the complex plane, care must
be taken that its values used in \eqref{eq:Sk} are taken from the same
sheath.

The eigenvalue problem \eqref{eq:evalda} needs to be restricted to
eigenfunctions {satisfying condition} \eqref{eq:evaldb} and this
is done with a projection approach. Defining $\u =
[\alpha_0,\dots,\alpha_{N-1}]^T$ and the matrix $\bM := - i\, \bA^{-1}
(\bB + \bC + \bD )$, equation \eqref{eq:evalda} can be expressed as
\begin{equation}
\lambda \, \u = \bM\, \u.
\label{eq:M}
\end{equation}
Introducing operator $\b \; : \; \RR^N \longrightarrow \RR$ defined as
\begin{equation}
\b = {\left[2,\ 0, \ -\frac{2}{3}, \ 0, \ -\frac{2}{15},\ \dots,
\ -\frac{2}{(N-1)^2 - 1}, \ 0 \right]},
\label{eq:b}
\end{equation} 
the constraint \eqref{eq:evaldb} can be expressed as $\b\u = 0$. The
kernel space of this operator, $\N(\b)$, thus corresponds to the
subspace of functions {satisfying condition \eqref{eq:evaldb}}.
The projection onto this subspace is realized by the operator
$\P_{\N(\b)} := \mathbf{I} - \b^{\dagger} \b$, where $\mathbf{I}$ is
the $N\times N$ identity matrix and $\b^{\dagger} := \b^T
(\b\b^T)^{-1} = {\b^T / ||\b||_2}$ is the Moore-Penrose
pseudo-inverse of the operator $\b$ \citep{l05}.  Defining the
restricted variable $\z := \P_{\N(\b)} \, \u$, problem
\eqref{eq:M} transforms to
\begin{equation}
\lambda \, \z = \P_{\N(\b)}\, \bM \, \P^{\dagger}_{\N(\b)} \,\z =: \bM_{\b} \, \z,
\label{eq:Mb}
\end{equation}
where $\P^{\dagger}_{\N(\b)}$ is the Moore-Penrose pseudo-inverse of
the projector $\P_{\N(\b)}$, which can now be solved using standard
techniques. We note that the dimension of this problem is $N-1$. An
alternative approach to impose constraint {\eqref{eq:evaldb} is}
to frame \eqref{eq:evald} as a generalized eigenvalue problem
\citep{l05}.

We now offer some comments about the accuracy and validation of the
computational approach described above. The accuracy of approximation
of singular integrals in \eqref{eq:intI2} and \eqref{eq:intI1} was
tested by applying this approach to the integral operator in
\eqref{eq:vn} which has the same singularity structure as $\T(\theta)$
in \eqref{eq:intI2} and $\T_k(\theta)$ in \eqref{eq:intI1}, and for
which an exact formula is available, cf.~\eqref{eq:psiHill}. In
addition, an analogous test was conducted for the tangential velocity
component given by $\C \, \t_{\theta}\cdot\left[ \int_0^{\pi}
  \bK(\theta,\theta')\, d\theta' - W\, \e_x\right]$. Using {\tt
  maxLength}=$10^6$ (which controls the length of the Chebyshev series
in {\tt chebfun}) resulted in $L_\infty$ errors of order
$\O(10^{-12})$ which is close to the machine precision. The rather
complicated analytical expression \eqref{eq:Sk} used in
$\mathcal{S}_k(\theta)$, involving multi-valued functions with branch
cuts in the complex plane, was validated by comparing it against a
numerical approximation of the weakly-singular integral defining
$\mathcal{S}_k(\theta)$.  With the high precision of the numerical
quadratures thus verified, the shape differentiation results in
\eqref{eq:vne_rhs} were validated by comparing them against simple
forward finite-difference approximations of the shape derivatives.
For example, the consistency of the first term on the RHS in
\eqref{eq:vne_rhs} was checked by comparing it (as a function of
$\theta$) to
\begin{equation*}
\epsilon^{-1} \, \C \, \left( \n_{\x^{\epsilon}} - \n_{\x} \right) \cdot 
\left[ \int_{\partial\A} \bK(\x(t),\x')\, ds_{\x'} - W\, \e_x\right]
\end{equation*}
in the limit of vanishing $\epsilon$. In the same spirit, the
consistency of the second and third term on the RHS of
\eqref{eq:vne_rhs} was verified by perturbing the evaluation point
$\x$ and the contour $\partial\A$, respectively. We also checked
computationally that the projection formulation \eqref{eq:Mb} of the
constrained eigenvalue problem \eqref{eq:evalp} gives
{essentially} the same results as its formulation in terms of a
generalized eigenvalue problem {(the former approach was in fact
  found to be somewhat more sensitive to round-off errors owing to the
  conditioning of the projection operator $\P_{\N(\b)}$)}. The
algebraic eigenvalue problem was solved in {\tt Matlab} with the QR
and Cholesky methods producing essentially identical results.

Anticipating the results of \S \ref{sec:results}, we now introduce a
regularized version of eigenvalue problem \eqref{eq:Mb} in which it is
ensured that the coefficients $\alpha_k$ decay with the wavenumber $k$
sufficiently rapidly, thus guaranteeing the required regularity of the
eigenvectors $u$. We introduce the following diagonal operator acting
as a low-pass filter
\begin{equation}
F_{jk}^{\delta} := \left\{
\begin{aligned}
& \frac{1}{1+\left(\delta \, k\right)^{2p}}, & \quad & j=k, \\
& 0, & & j \neq k,
\end{aligned}\right.
\label{eq:F}
\end{equation}
where $\delta>0$ is the cut-off length scale and $p$ a positive
integer, and define $\z^{\delta} := \bF^{\delta} \, \z$. This filter
can be regarded as the discretization of the elliptic operator
$\left[\operatorname{Id} - (-1)^{p+1}\delta^{2p} \,
  \Dpartialn{}{\theta}{2p}\right]^{-1}$ in the trigonometric basis. We
then obtain from \eqref{eq:Mb}
\begin{equation}
\lambda \, \z^{\delta} 
= \bF^{\delta} \, \bM_{\b} \,\left(\bF^{\delta}\right)^{-1} \, \z^{\delta}
=: \bM_{\b}^{\delta} \,\z^{\delta},
\label{eq:Md}
\end{equation}
from which the original problem is clearly recovered when $\delta
\rightarrow 0$. The regularized eigenvectors $u^{\delta}$
corresponding to $\z^{\delta}$ are therefore guaranteed to be smoother
than the original eigenvectors $u$ (the actual improvement of
smoothness depends on the value of $p$). In the next section, among
other results, we will study the behavior of solutions to eigenvalue
problem \eqref{eq:Md} for a decreasing sequence of regularization
parameters $\delta$.

\section{Computational Results}
\label{sec:results}

In this section we first summarize the numerical parameters used in
the computations and then present the results obtained by solving
eigenvalue problem \eqref{eq:Md} for different values of the
regularization parameter $\delta$. All computations were conducted
setting $p=4$ in the regularization operator \eqref{eq:F} and using
the resolutions $N=64,128,256,512,1024$ in \eqref{eq:un}. We allowed
the regularization parameter to take a wide range of values $\delta
=1, \frac{1}{2}, \frac{1}{4},\dots, \frac{1}{1024}$. We note that with
the smallest values from this range regularization barely affects the
eigenvalue problem \eqref{eq:Md} even when the highest resolutions are
used.  Therefore, these value may be considered small enough to
effectively correspond to the limit $\delta \rightarrow 0$.

\begin{figure}
\begin{center}
\includegraphics[width=0.8\textwidth]{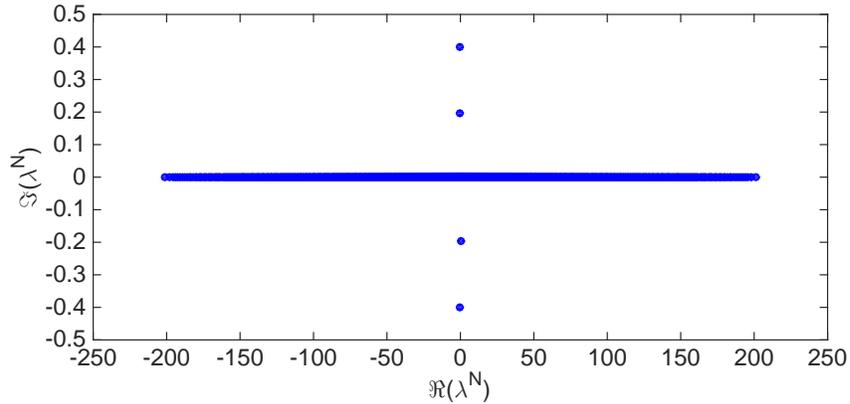}
\end{center}
\caption{Eigenvalue spectrum of problem \eqref{eq:Md} obtained with
  $N=1024$ and $\delta = \frac{1}{32}$.}
\label{fig:sp}
\end{figure}

In our analysis below we will first demonstrate that, for a fixed
parameter $\delta$, the solutions of the regularized problem
\eqref{eq:Md} converge as the numerical resolution $N$ is refined.
Then, we will study the behavior of the eigenvalues and eigenfunctions
as the regularization parameter $\delta$ is reduced.  A typical
eigenvalue spectrum obtained by solving problem \eqref{eq:Md} is shown
in figure \ref{fig:sp}. The fact that the spectrum is symmetric with
respect to the lines $\Re(\lambda) = 0$ and $\Im(\lambda) = 0$
reflects the Hamiltonian structure of the problem. Given the ansatz
for the perturbations introduced in \S \ref{sec:stab}, eigenvalues
with negative imaginary parts correspond to linearly unstable
eigenmodes and we see in figure \ref{fig:sp} that there are two such
eigenvalues in addition to two eigenvalues associated with linearly
stable eigenmodes. We will refer as the first and the second to the
eigenvalues with, respectively, the larger and the smaller magnitude.
In addition to these four purely imaginary eigenvalues, there is also
a large number of purely real eigenvalues covering a segment of the
axis $\Im(\lambda) = 0$ which can be interpreted as the {\em
  continuous} spectrum. The spectrum shown in figure \ref{fig:sp} was
found to be essentially independent of the regularization parameter
$\delta$ and its dependence on the numerical resolution $N$ is
discussed below. In this analysis we will set $\delta = \frac{1}{32}$.

\begin{figure}
\begin{center}
\mbox{
\subfigure[]{\includegraphics[width=0.5\textwidth]{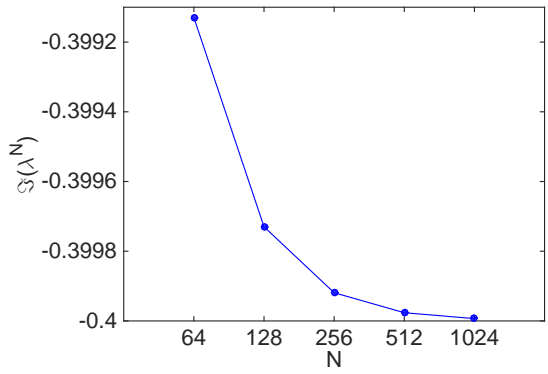}}\quad
\subfigure[]{\includegraphics[width=0.5\textwidth]{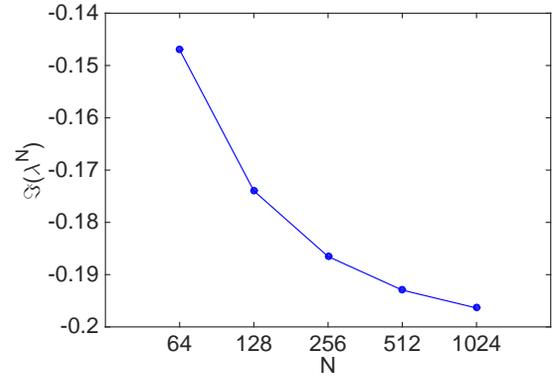}}
}
\mbox{
\subfigure[]{\includegraphics[width=0.5\textwidth]{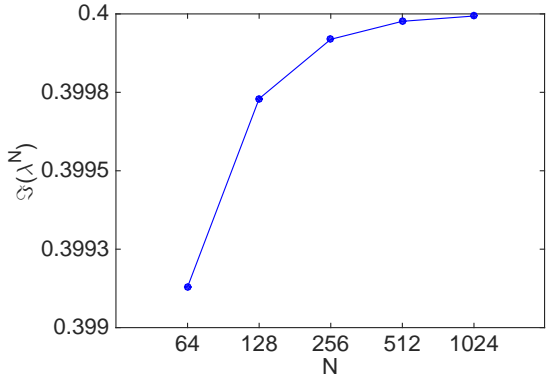}}\quad
\subfigure[]{\includegraphics[width=0.5\textwidth]{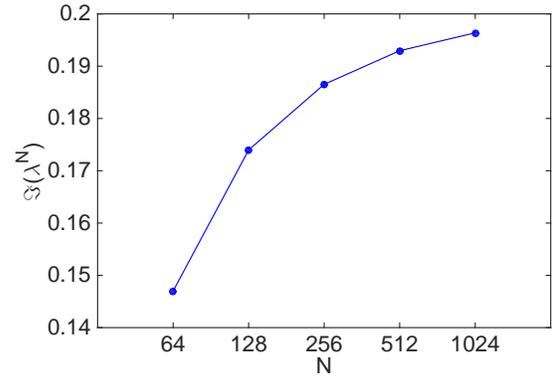}}
}
\end{center}
\caption{Dependence of the four purely imaginary eigenvalues on the
  resolution $N$ with a fixed regularization parameter $\delta =
  \frac{1}{32}$. The eigenvalues shown in (a) and (b) correspond to
  the two unstable eigenmodes, {whereas those shown in (c) and
    (d) correspond to the two stable eigenmodes}.}
\label{fig:ImLam}
\end{figure}

\begin{figure}
\begin{center}
\mbox{
\subfigure[]{\includegraphics[width=0.5\textwidth]{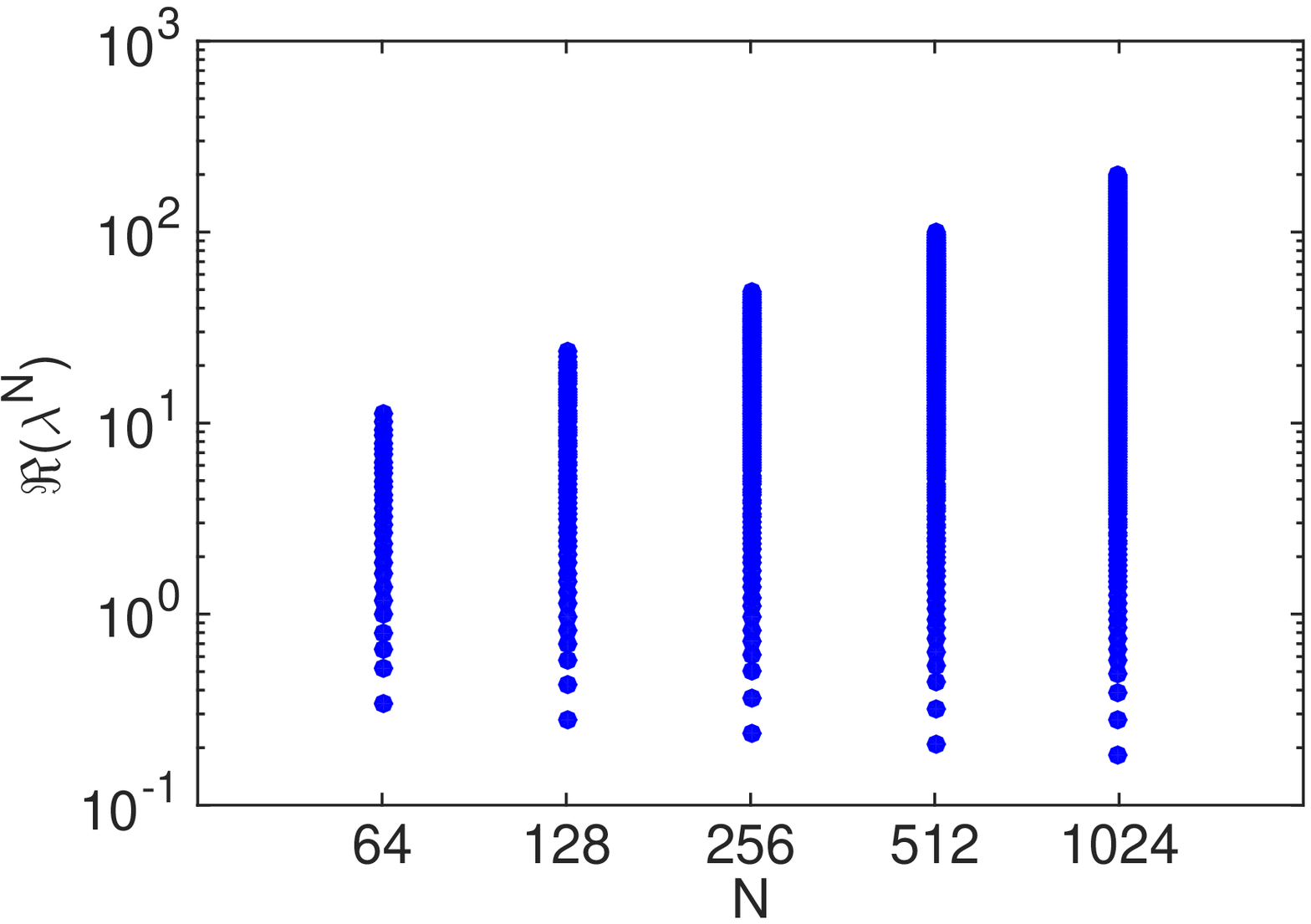}}\quad
\subfigure[]{\includegraphics[width=0.5\textwidth]{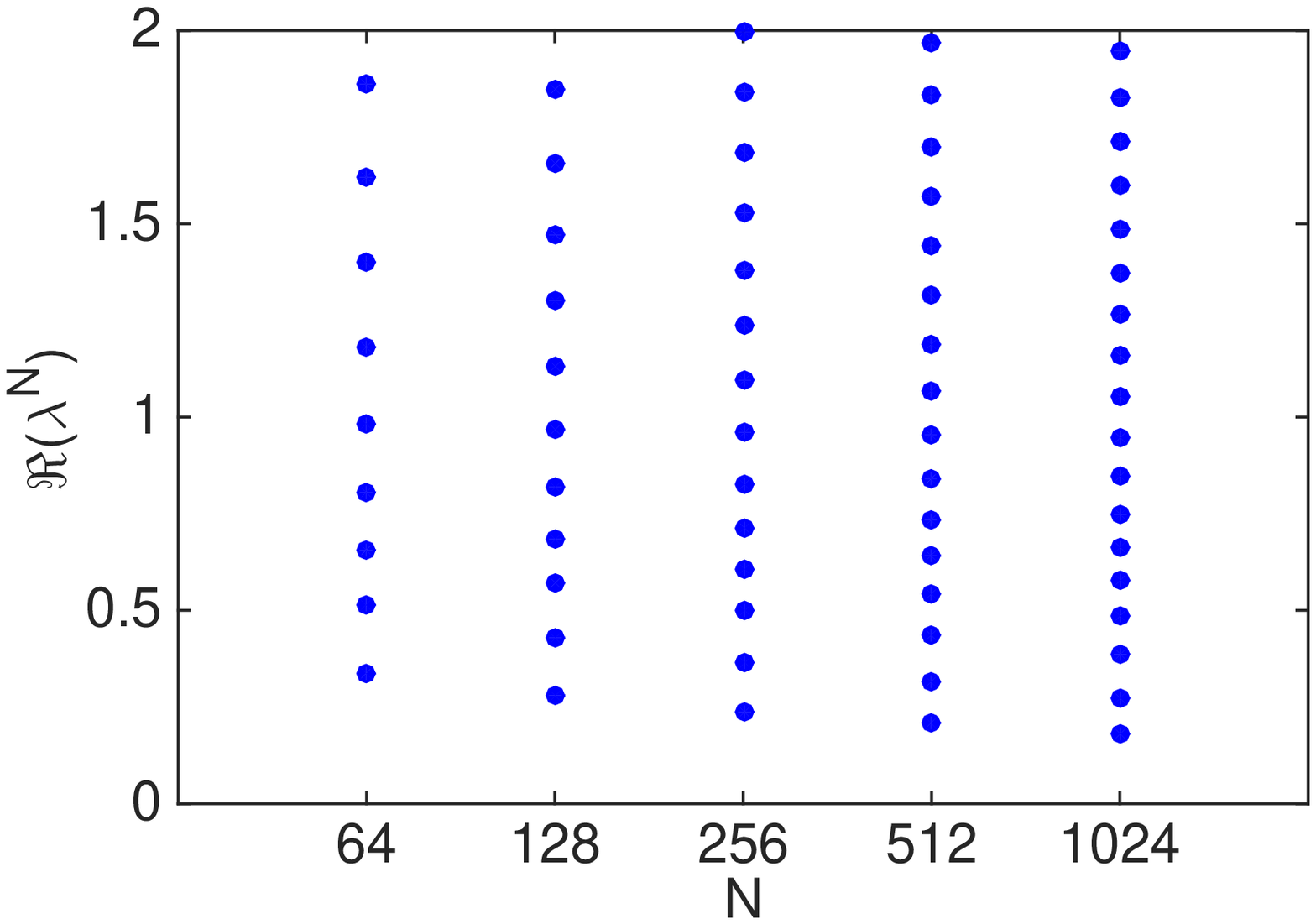}}
}
\end{center}
\caption{Dependence of the purely real eigenvalues on the resolution
  $N$ with a fixed regularization parameter $\delta = \frac{1}{32}$
  plotted (a) using the logarithmic scale and (b) within the interval
  $[0,2]$ using the linear scale for the vertical axis. For clarity,
  only the eigenvalues with positive real parts ($\Re(\lambda_i^N) >
  0$) are shown.}
\label{fig:ReLam}
\end{figure}

The dependence of the four purely imaginary eigenvalues on the
resolution $N$ is shown in figures \ref{fig:ImLam}(a--d), where we see
that the eigenvalues all converge to well-defined limits.
{However, as will be discussed below, problem \eqref{eq:evald}
  does not admit smooth solutions (eigenvectors) and therefore the
  convergence of eigenvalues $\lambda^N$ with $N$ is only algebraic
  rather than spectral. Thus, the numerical approximation error for an
  eigenvalue $\lambda$ can be represented as $|\lambda^N - \lambda| =
  c N^{\beta}$ for some $c>0$ and $\beta<0$. Using the data from
  figure \ref{fig:ImLam} to evaluate $(\lambda^N - \lambda^{2N})$ as a
  function of the resolution $N$, one can estimate the order of
  convergence using a least-squares fit as $\beta \approx -1.72$ for
  the first eigenvalue (both stable and unstable) and $\beta \approx
  -0.99$ for the second (both stable and unstable). This confirms that
  the first eigenvalues converge much faster than the second.} The
dependence of the purely real eigenvalues on the resolution $N$ is
illustrated in figures \ref{fig:ReLam}(a--b). First of all, we notice
that the purely real eigenvalues do not appear to converge to any
particular limit as $N$ is increased and instead fill the interval of
the axis $\Im(\lambda)=0$ with increasing density (figure
\ref{fig:ReLam}(b)).  From figure \ref{fig:ReLam}(a) we can infer that
the lower and upper bounds of this interval approximately scale with
the resolution as
\begin{equation}
\min_i \left[|\Re(\lambda_i^N)|\right] \sim N^{-0.22} 
\quad \text{and} \quad 
\max_i \left[|\Re(\lambda_i^N)|\right] \sim N^{1.04}, \quad i=1,\dots,N,
\label{eq:lm}
\end{equation}
where $\lambda_i^N$ denotes the $i$-th eigenvalue computed with the
resolution $N$. All these observations allow us to conclude that the
continuous eigenvalue problem \eqref{eq:evalp} has four purely
imaginary eigenvalues and a continuous spectrum coinciding with the
axis $\Im(\lambda) = 0$.

\begin{figure}
\begin{center}
\mbox{
\subfigure[]{\includegraphics[width=0.5\textwidth]{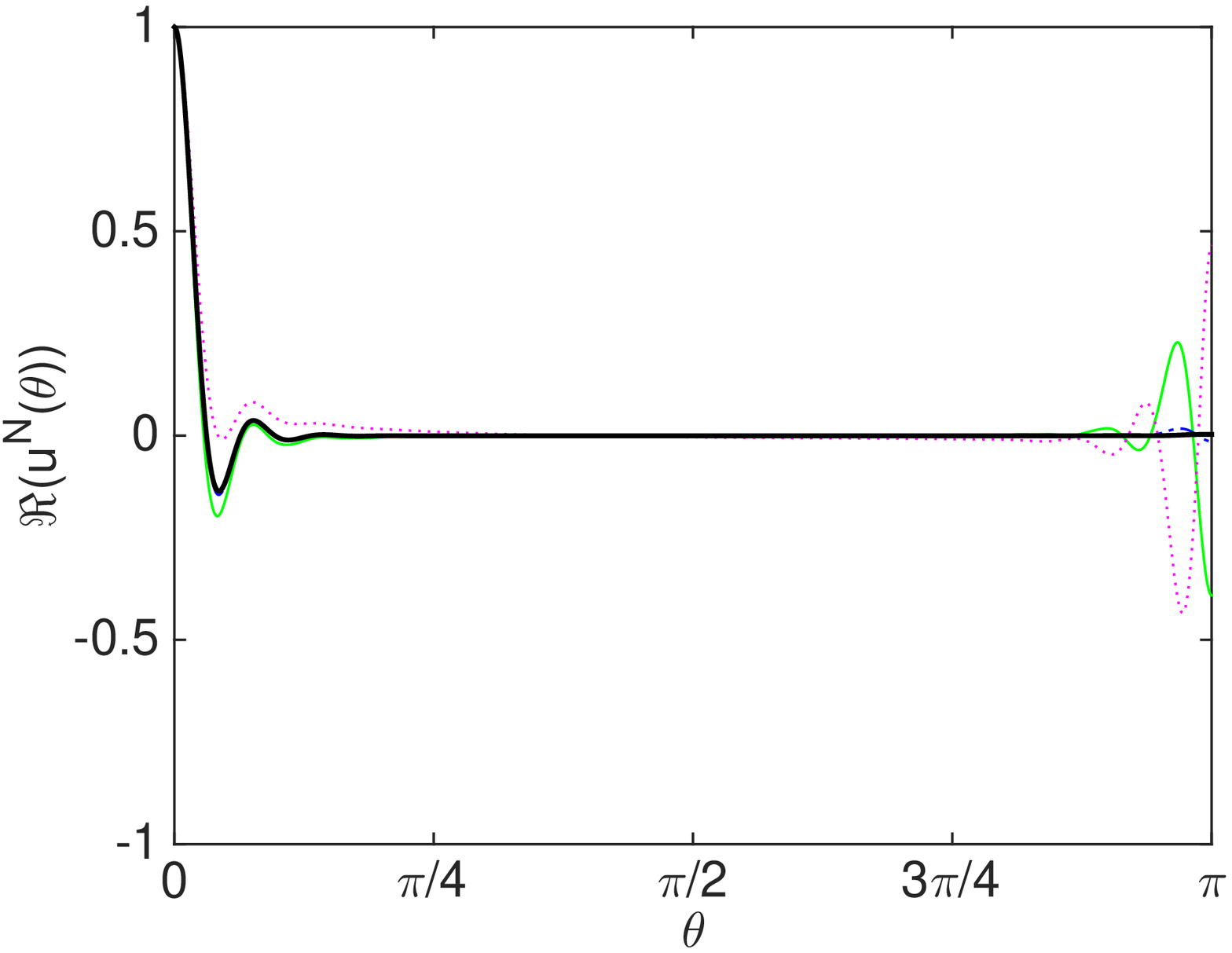}}\quad
\subfigure[]{\includegraphics[width=0.5\textwidth]{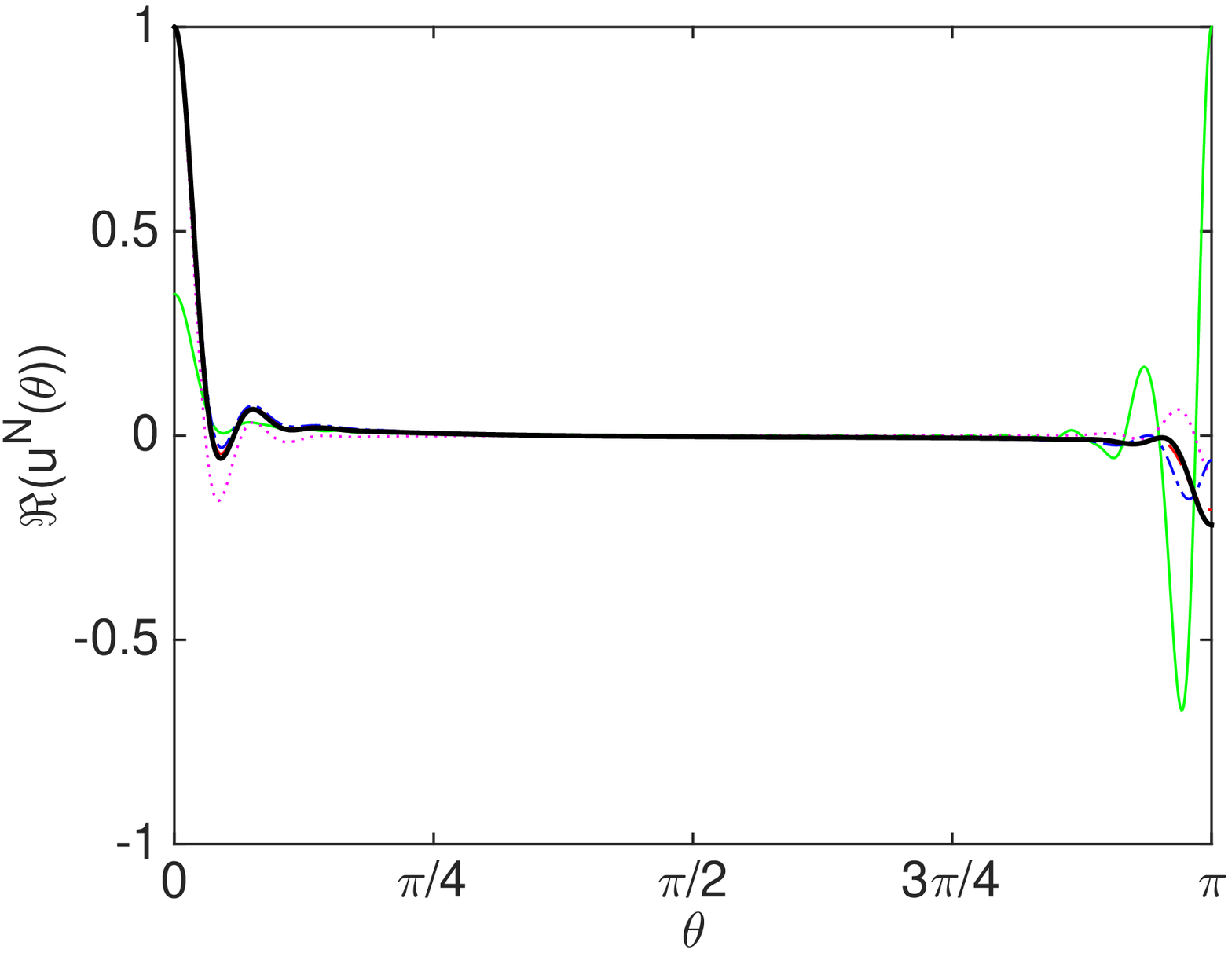}}
}
\mbox{
\subfigure[]{\includegraphics[width=0.5\textwidth]{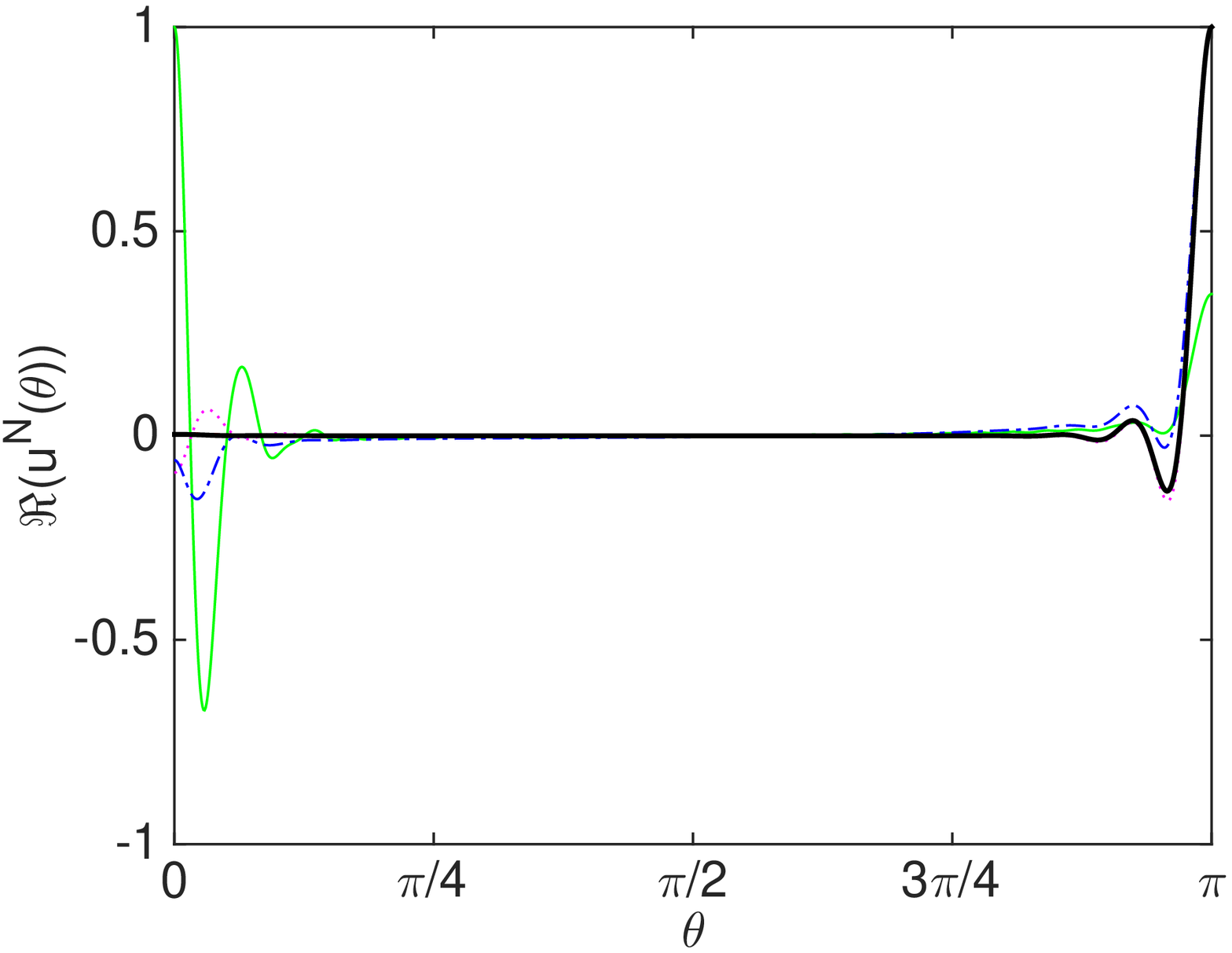}}\quad
\subfigure[]{\includegraphics[width=0.5\textwidth]{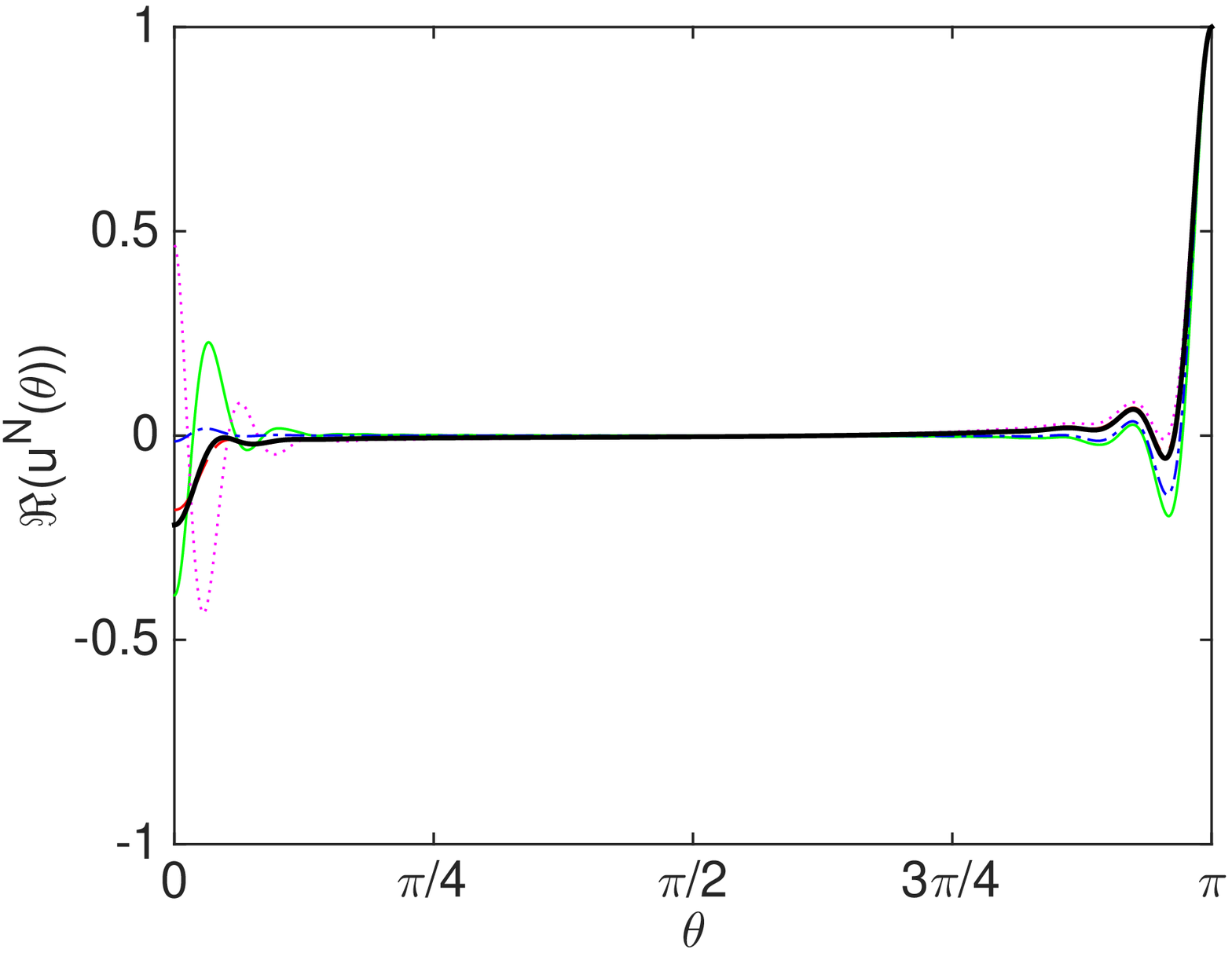}}
}
\end{center}
\caption{Unstable eigenvectors corresponding to the first (a) and
  second (b) eigenvalue, and stable eigenvectors corresponding to the
  first (c) and second (d) eigenvalue for different resolutions:
  $N=64$ (green solid line), $N=128$ (magenta dotted line), $N=256$
  (blue dash-dotted line), $N=512$ (red dashed line), $N=1024$ (thick
  black solid line) with $\delta = \frac{1}{32}$. {The
    eigenvectors are normalized such that $\sup_{\theta \in [0,\pi]}
      |\Re(u^N(\theta))| = 1$.}}
\label{fig:evec_N_th}
\end{figure}

\begin{figure}
\begin{center}
\mbox{
\subfigure[]{\includegraphics[width=0.5\textwidth]{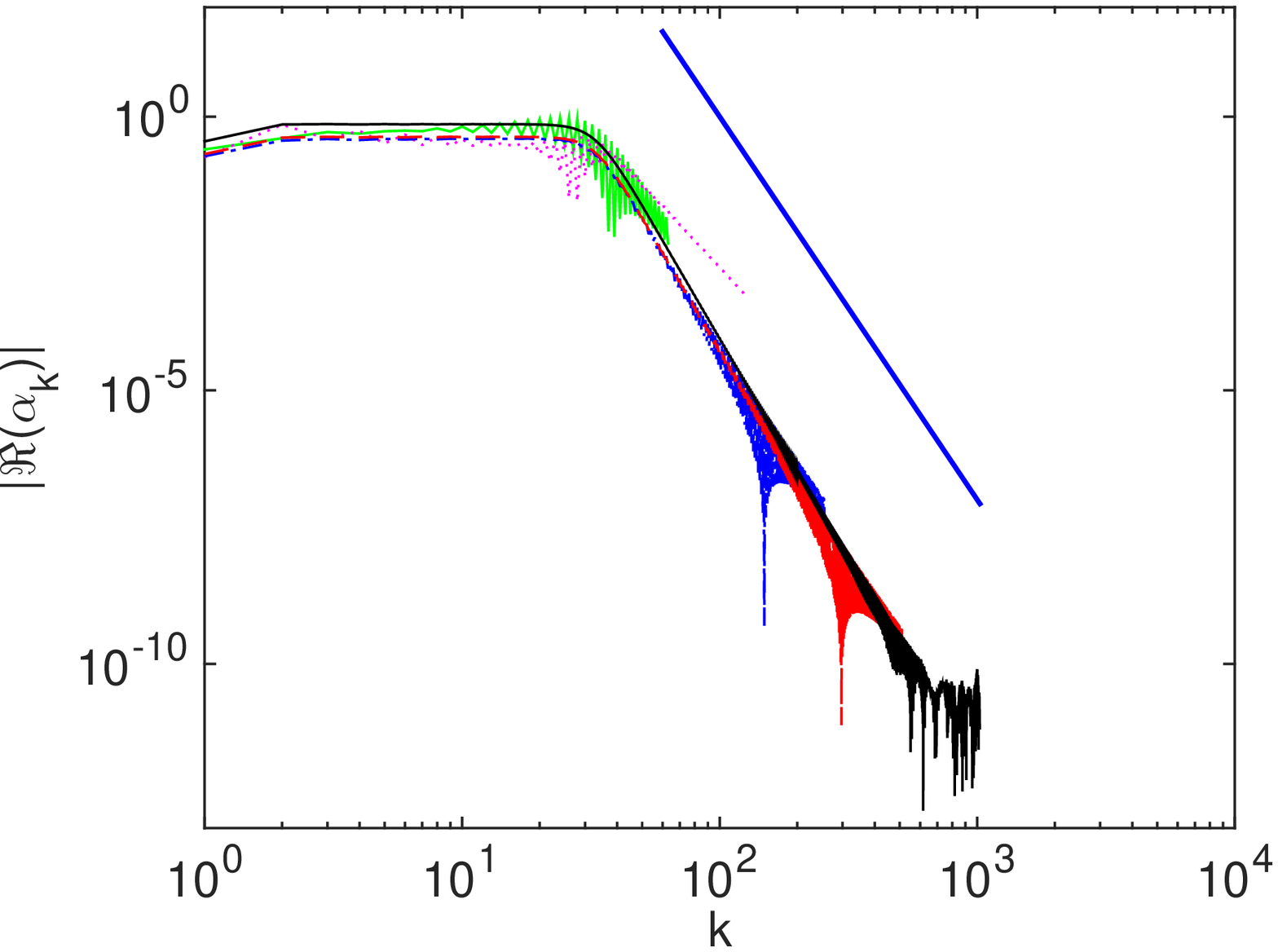}}\quad
\subfigure[]{\includegraphics[width=0.5\textwidth]{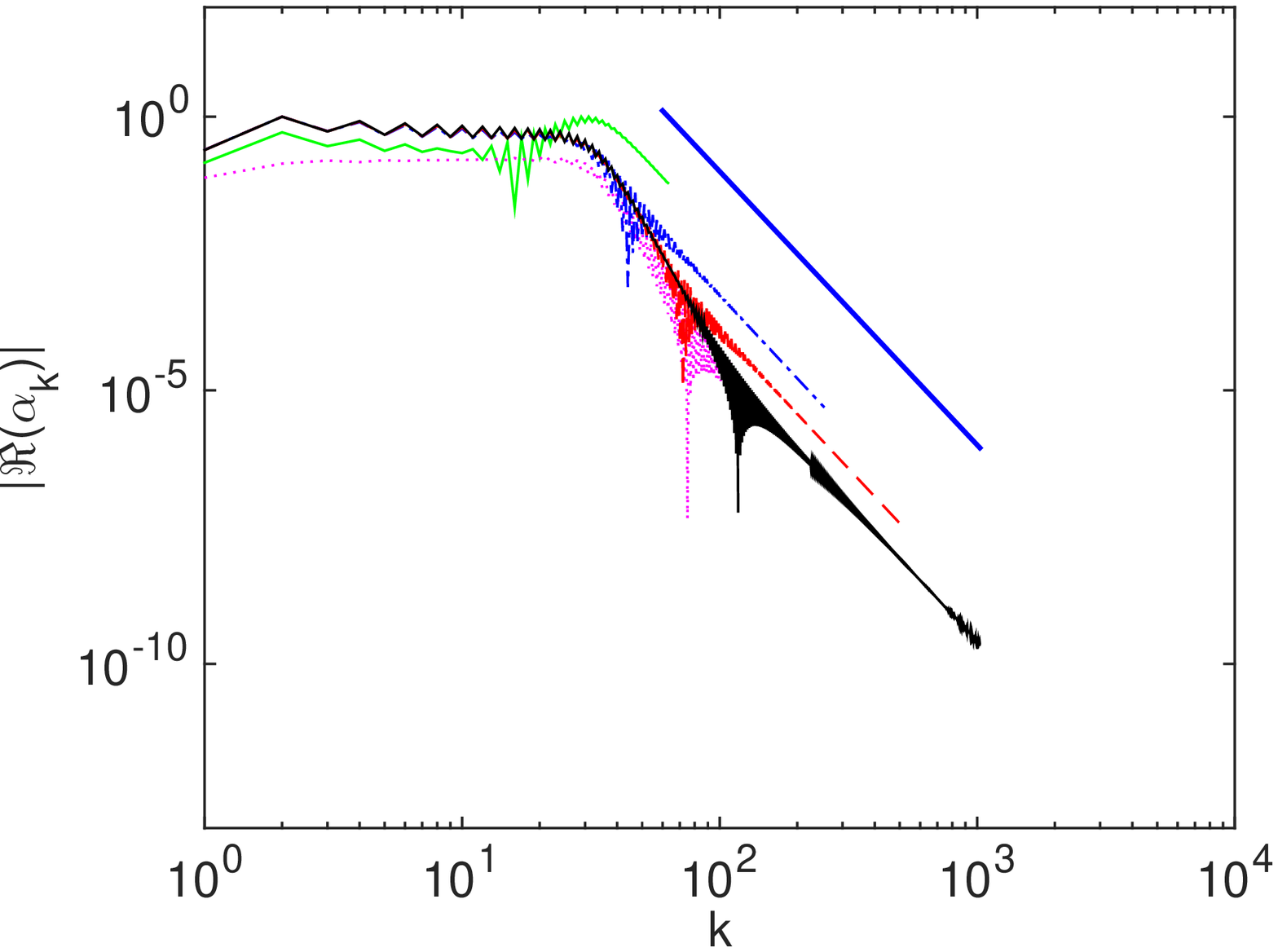}}
}
\end{center}
\caption{Fourier coefficient spectra ${|\Re(\alpha_k)|}$, $k=1,\dots,N$,
  characterizing the eigenvectors associated with the first (a) and
  the second (b) eigenvalue for different resolutions: $N=64$ (green
  solid line), $N=128$ (magenta dotted line), $N=256$ (blue
  dash-dotted line), $N=512$ (red dashed line), $N=1024$ (thick black
  solid line).  The spectra of the stable and unstable eigenvectors
  corresponding to eigenvalues with the same magnitude are identical.
  The straight blue solid lines have the slopes of {$-7$} (a) and
  {$-5$} (b).}
\label{fig:evec_N_sp}
\end{figure}

We now move on to discuss the eigenvectors corresponding to the purely
imaginary eigenvalues. The linearly unstable and stable eigenvectors
are shown as functions of the polar angle $\theta$ for different
resolutions $N$ in figures \ref{fig:evec_N_th}(a--d). {In these
  figures we only show the real parts of the eigenvectors, since given
  our ansatz \eqref{eq:r} for the perturbation, the imaginary parts do
  not play any role when the eigenvalues are purely imaginary. Hence,
  below the term ``eigenvector'' will refer to $\Re(u^N(\theta))$.}
We note that, {as the resolution $N$ increases,} the unstable and
stable eigenvectors associated with a given eigenvalue {become}
reflections of each other with respect to the midpoint $\theta = \pi /
2$ with the unstable eigenvectors exhibiting a localized peak near the
rear stagnation point ($\theta = 0$) and the stable eigenvectors
exhibiting such a peak near the front stagnation point ($\theta =
\pi$). In figures \ref{fig:evec_N_th}(a--d) we also observe that, for
a fixed regularization parameter $\delta$, the numerical
approximations {$\Re(u^N(\theta))$} of eigenfunctions converge
uniformly in $\theta$ for increasing $N$, although this convergence is
significantly slower for points $\theta$ close to the endpoint
opposite to where the eigenvector exhibits a peak. We remark that the
same behaviour of spectral approximations to eigenfunctions was also
observed by \citet{r99}. {The two unstable eigenvectors
  $\Re(u^N_1)$ and $\Re(u^N_2)$ are strongly non-normal with $\big\langle
    \Re(u^N_1), \Re(u^N_2) \big\rangle_{L_2} / \left( || \Re(u^N_1)
    ||_{L_2} || \Re(u^N_2) ||_{L_2} \right) \approx 0.96$, where
  $\langle\cdot,\cdot\rangle_{L_2}$ and $||\cdot||_{L_2}$ are respectively the
  inner product and the norm in the space $L_2(0,\pi)$, when $\delta =
  1/32$ and $N=1024$. Consequently,} the two unstable eigenvectors
appear quite similar as functions of $\theta$, especially near the
peak (cf.~figures \ref{fig:evec_N_th}(a--b)). {On the other
  hand,} the Fourier spectra of their expansion coefficients shown in
figures \ref{fig:evec_N_sp}(a--b) {exhibit quite distinct
  properties.} More specifically, we see that the slope of the Fourier
spectra for $k > 1/\delta$ is quite different in the two cases: it is
{close to $-7$ and $-5$} for the eigenvectors associated with,
respectively, the first and second eigenvalue.  We emphasize however
that the specific slopes are determined by the choice of the parameter
$p$ in the regularizing operator \eqref{eq:F} and here we are
interested in the relative difference of the slopes in the two cases.
Further distinctions between the eigenvectors associated with the
first and the second eigenvalue will be elucidated below when
discussing their behavior in the limit of decreasing regularization
parameter $\delta$.

\begin{figure}
\begin{center}
\mbox{
\subfigure[]{\includegraphics[width=0.5\textwidth]{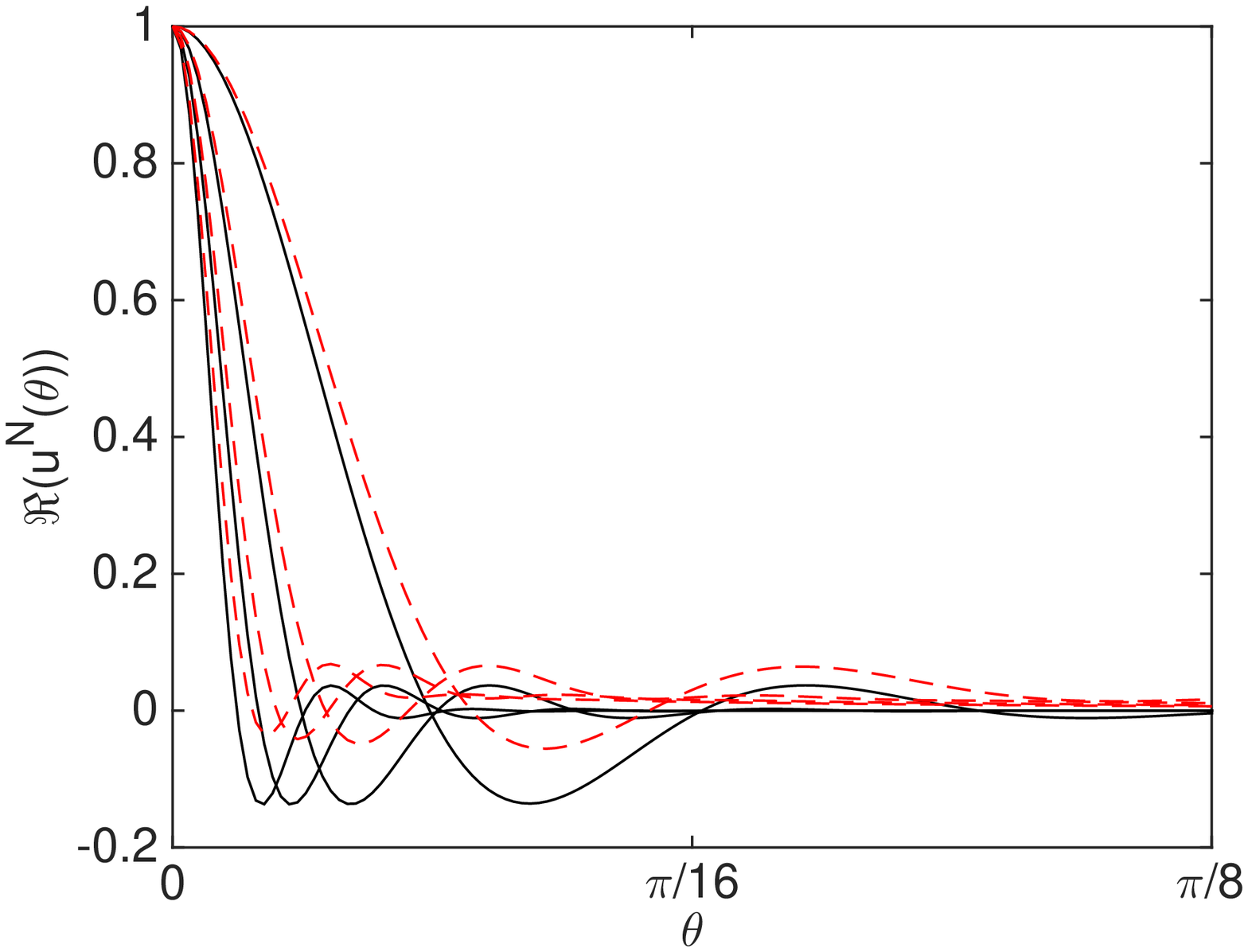}}\quad
\subfigure[]{\includegraphics[width=0.5\textwidth]{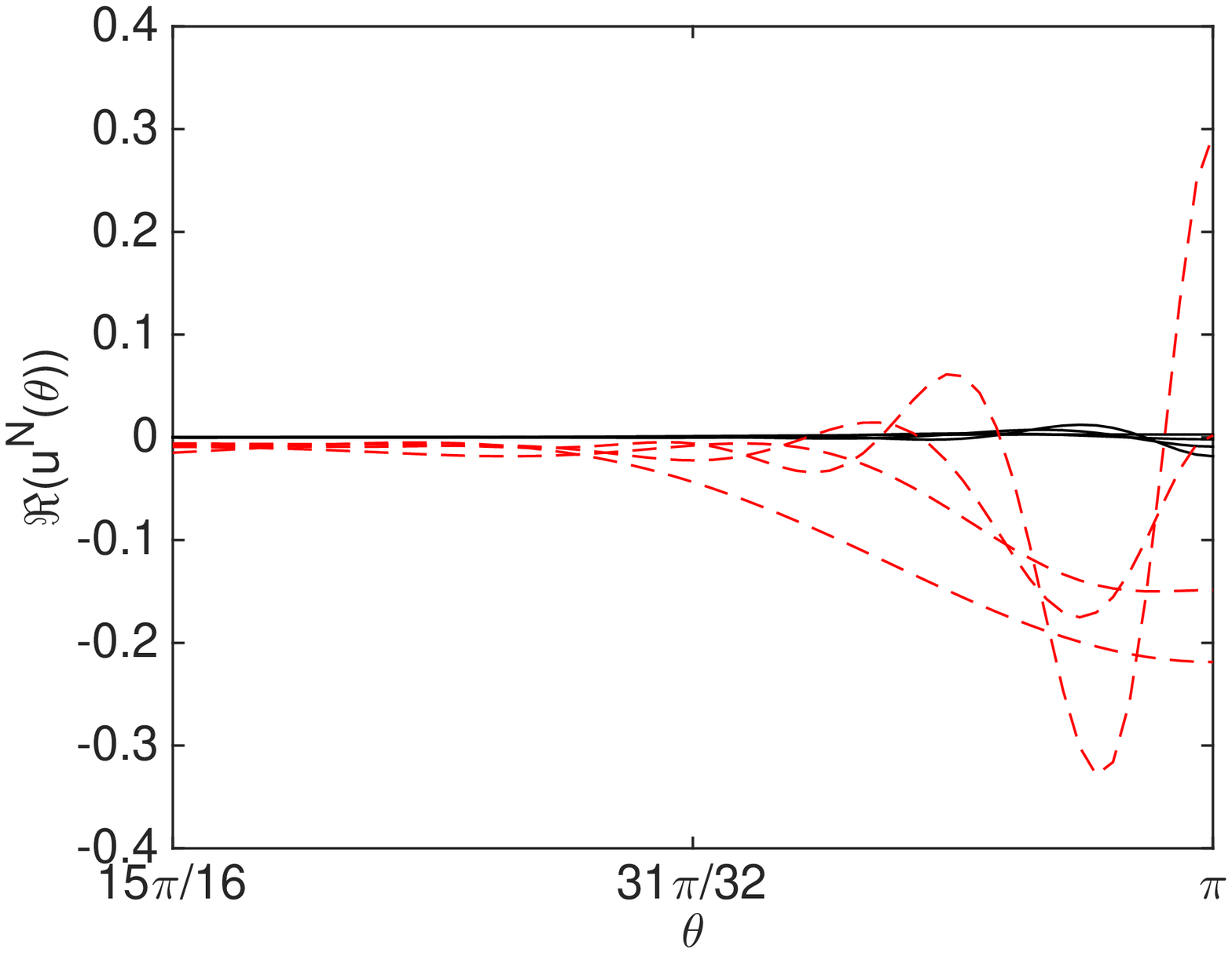}}
}
\end{center}
\caption{Dependence of the unstable eigenvectors associated with the
  first (black solid line) and the second (red dashed line) eigenvalue
  on the polar angle $\theta$ near the rear (a) and the front (b)
  stagnation point for different values of the regularization
  parameter $\delta = \frac{1}{32}, \frac{1}{64}, \frac{1}{96},
  \frac{1}{128}$ with $N=1024$ (smaller values of $\delta$ correspond
  to more localized eigenvectors). {The eigenvectors are
    normalized such that $\sup_{\theta \in [0,\pi]} |\Re(u^N(\theta))|
      = 1$.}}
\label{fig:evec_kc_th}
\end{figure}

\begin{figure}
\begin{center}
\mbox{
\subfigure[]{\includegraphics[width=0.5\textwidth]{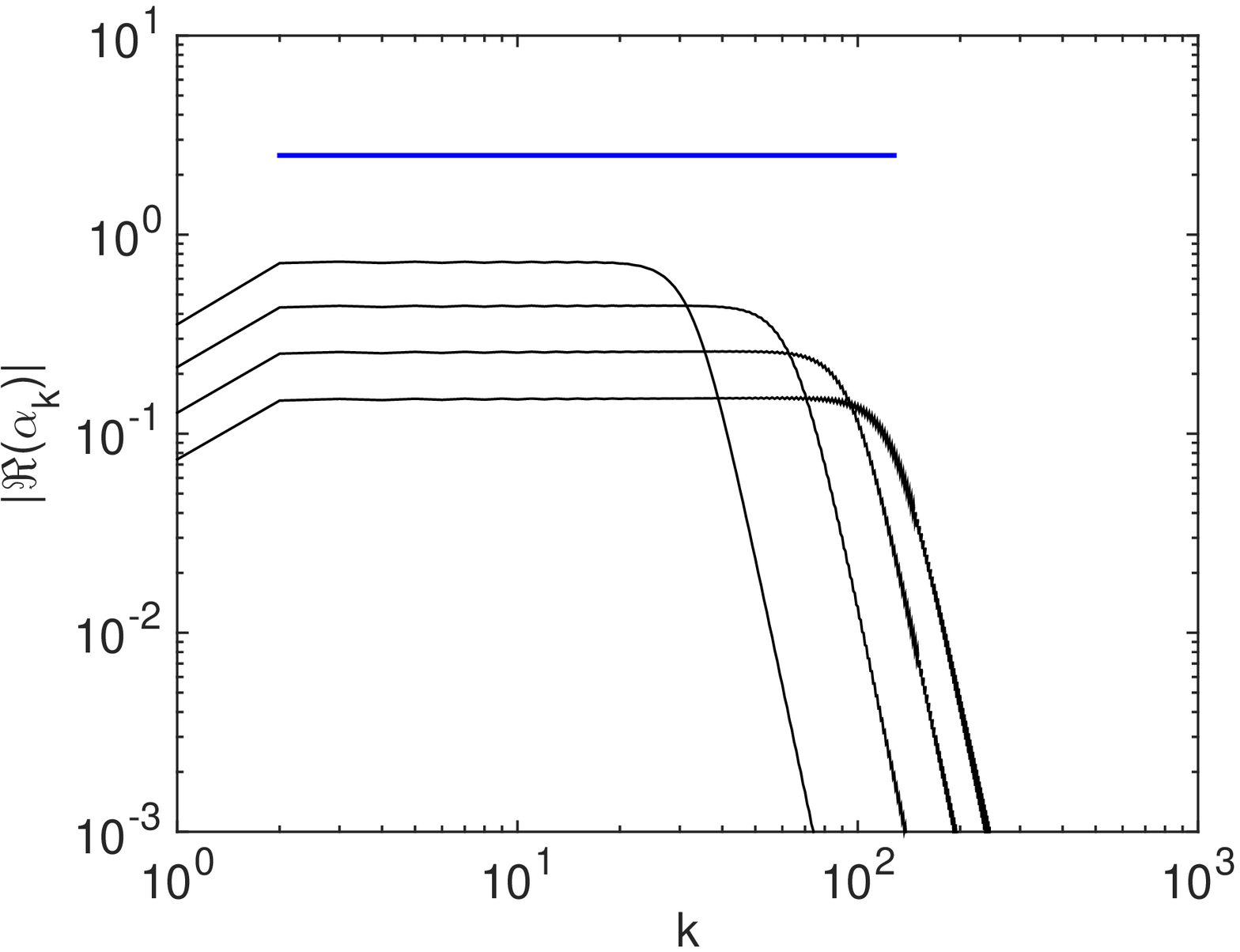}}\quad
\subfigure[]{\includegraphics[width=0.5\textwidth]{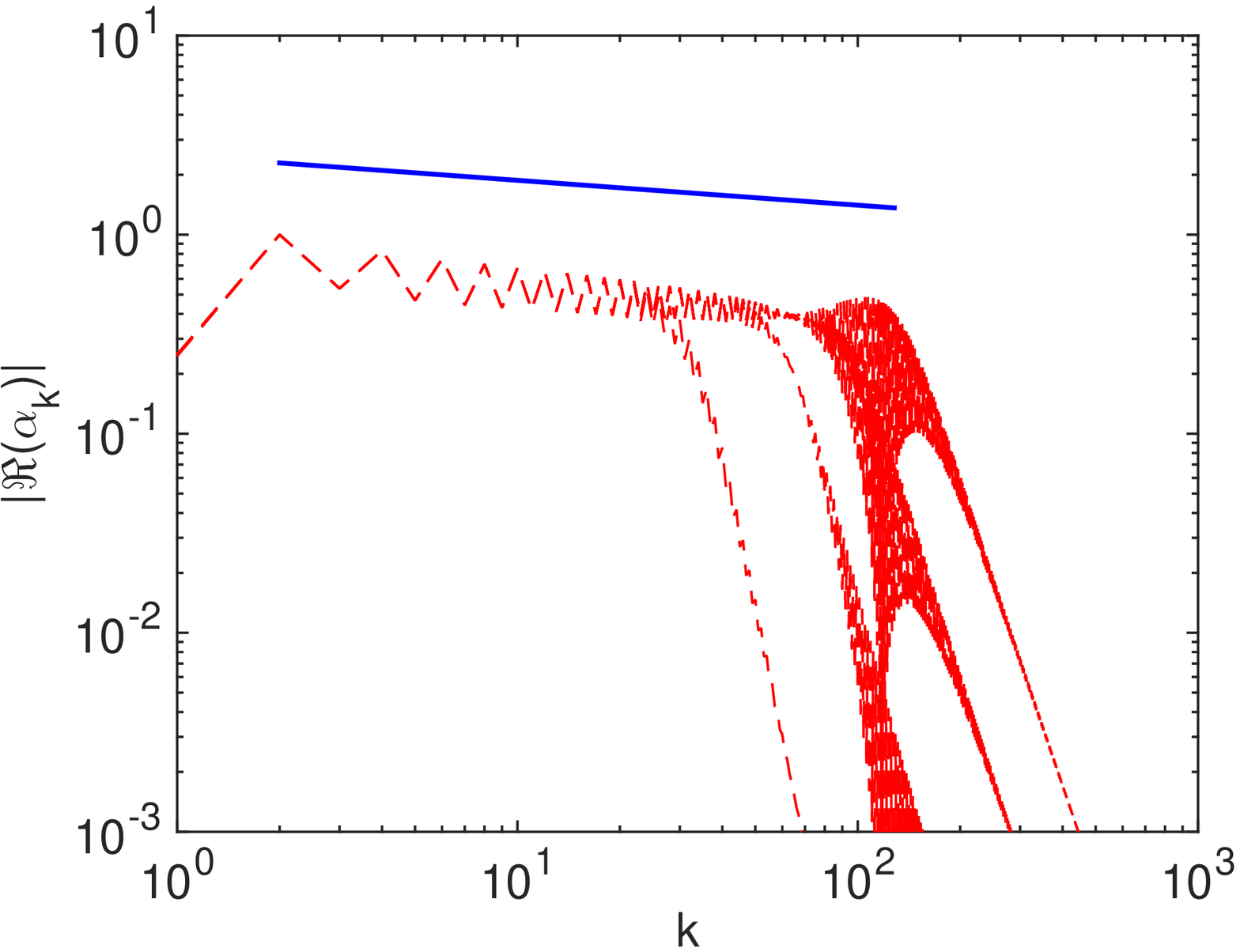}}
}
\end{center}
\caption{Fourier coefficient spectra ${|\Re(\alpha_k)|}$,
  $k=1,\dots,N$, of the eigenvectors associated with the first (a) and
  second (b) eigenvalue for different values of the regularization
  parameter $\delta = \frac{1}{32}, \frac{1}{64}, \frac{1}{96},
  \frac{1}{128}$ with $N=1024$ (smaller values of $\delta$ correspond
  to the plateau in ${|\Re(\alpha_k)|}$ extending to larger
  wavenumbers). The straight blue solid lines represent the slopes of
  {$0$} (a) and $-1/8$ (b).}
\label{fig:evec_kc_sp}
\end{figure}

Having established the convergence of the numerical approximations of
the eigenfunctions with the resolution $N$ for a fixed regularization
parameter $\delta$, we now go on to characterize their behaviour when
$\delta$ is decreased. Unless indicated otherwise, the results
presented below were obtained with the resolution $N=1024$. In figures
\ref{fig:evec_kc_th}(a--b) we show the behavior of the two unstable
eigenvectors near the rear and front stagnation points for different
values of the regularization parameter $\delta$. We see that, as this
parameter is decreased, the peak near the rear stagnation point
(figure \ref{fig:evec_kc_th}(a)) becomes steeper and more localized,
{especially for the eigenvector associated with the first
  eigenvalue}.  Likewise, the oscillation of the unstable eigenvectors
near the front stagnation point (figure \ref{fig:evec_kc_th}(b)) also
becomes more intense and localized as $\delta$ decreases, although
this effect is more pronounced in the case of the eigenvector
corresponding to the {second} eigenvalue. {These properties
  are further characterized} in the plots of the Fourier spectra of
the two eigenvectors shown in figures \ref{fig:evec_kc_sp}(a--b) for
different values of the regularization parameter $\delta$. In these
plots it is clear that, as the regularization effect vanishes
(corresponding to decreasing values of $\delta$), the point where the
slope of the spectrum changes moves towards larger wavenumbers $k$.
For $k < 1 / \delta$ the approximate slopes of the coefficient spectra
are, respectively, {$0$} and $-1/8$ for the eigenvectors
associated with the first and second eigenvalue (this difference of
slopes may explain the different behaviors in the physical space
already observed in figure \ref{fig:evec_kc_th}(b)).  Extrapolating
from the trends evident in figures \ref{fig:evec_kc_sp}(a--b) it can
be expected that these slopes will remain unchanged in the limit
$\delta\rightarrow 0$.  With this {behaviour} of the Fourier
coefficients, {involving no decay at all for the first
  eigenvector and a slow decay for the second,} expansion
\eqref{eq:un} does not converge in the limit of $N\rightarrow \infty$,
indicating that the stable and unstable eigenvectors do not have the
form of smooth functions, but rather ``distributions''. {As
  regards} the nature of their singularity, the slopes observed in
figures \ref{fig:evec_kc_sp}(a--b), {i.e., $0$ and $-1/8$,
  indicate that the eigenvector associated with the first eigenvalue
  is consistent with the Dirac delta (whose spectral slope is also 0),
  whereas the eigenvector associated with the second eigenvalue is
  intermediate} between the Dirac delta and the Heaviside step
function (whose spectral slope {is -1}).

\begin{figure}
\begin{center}
\includegraphics[width=0.5\textwidth]{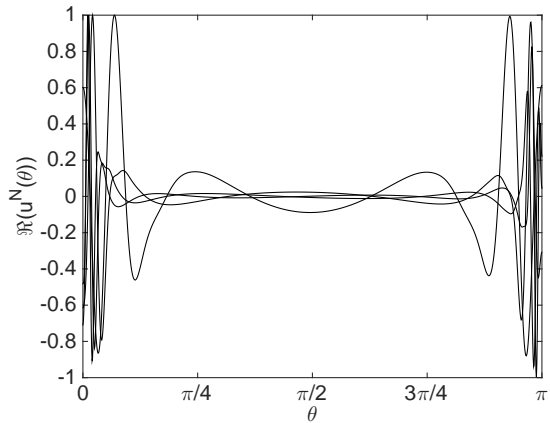}
\end{center}
\caption{Real part of the neutrally-stable eigenvector corresponding
  to the eigenvalue $\lambda \approx 1.0502$ computed with the
  resolution $N=1024$ and with different values of the regularization
  parameter $\delta = \frac{1}{32}, \frac{1}{64}, \frac{1}{96},
  \frac{1}{128}$ (smaller values of $\delta$ correspond to
  $\Re(u^N(\theta))$ exhibiting oscillations closer to the endpoints
  of the domain). {The eigenvectors are normalized such that
    $\sup_{\theta \in [0,\pi]} |\Re(u^N(\theta))| = 1$.}}
\label{fig:neut_th}
\end{figure}

\begin{figure}
\begin{center}
\mbox{
\subfigure[$\delta = \frac{1}{32}$]
{\includegraphics[width=0.5\textwidth]{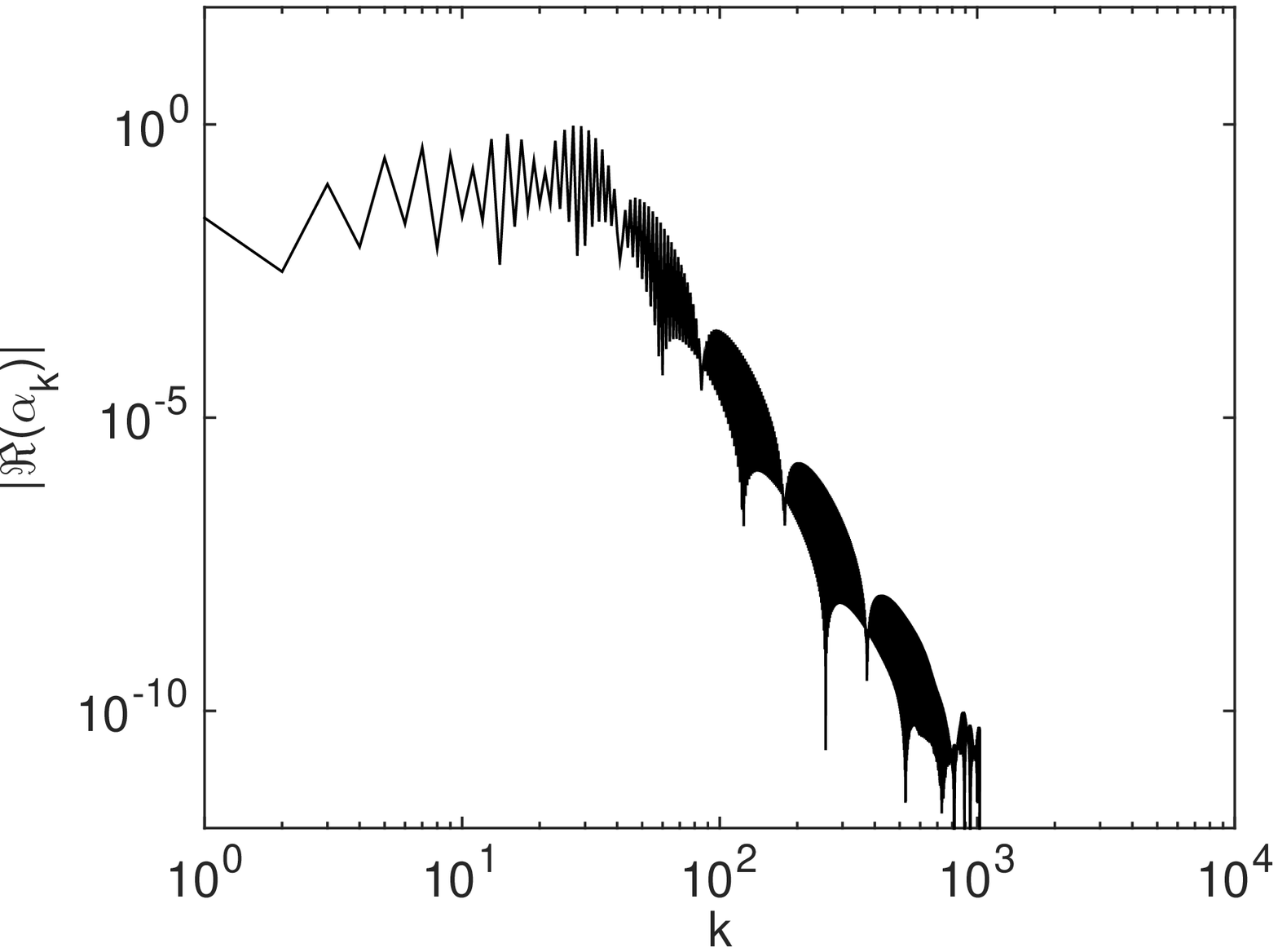}}\quad
\subfigure[$\delta = \frac{1}{64}$]
{\includegraphics[width=0.5\textwidth]{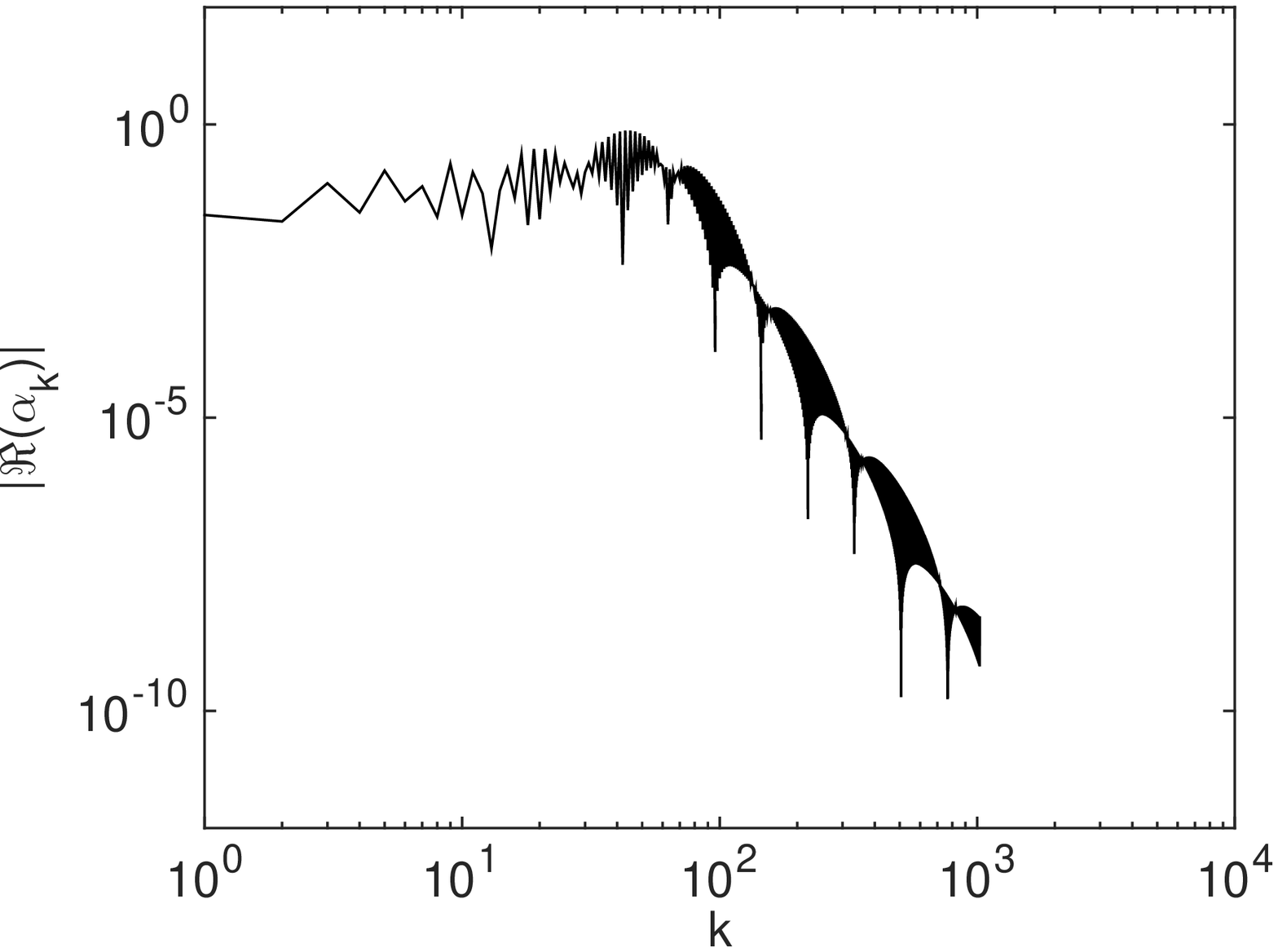}}
}
\mbox{
\subfigure[$\delta = \frac{1}{96}$]
{\includegraphics[width=0.5\textwidth]{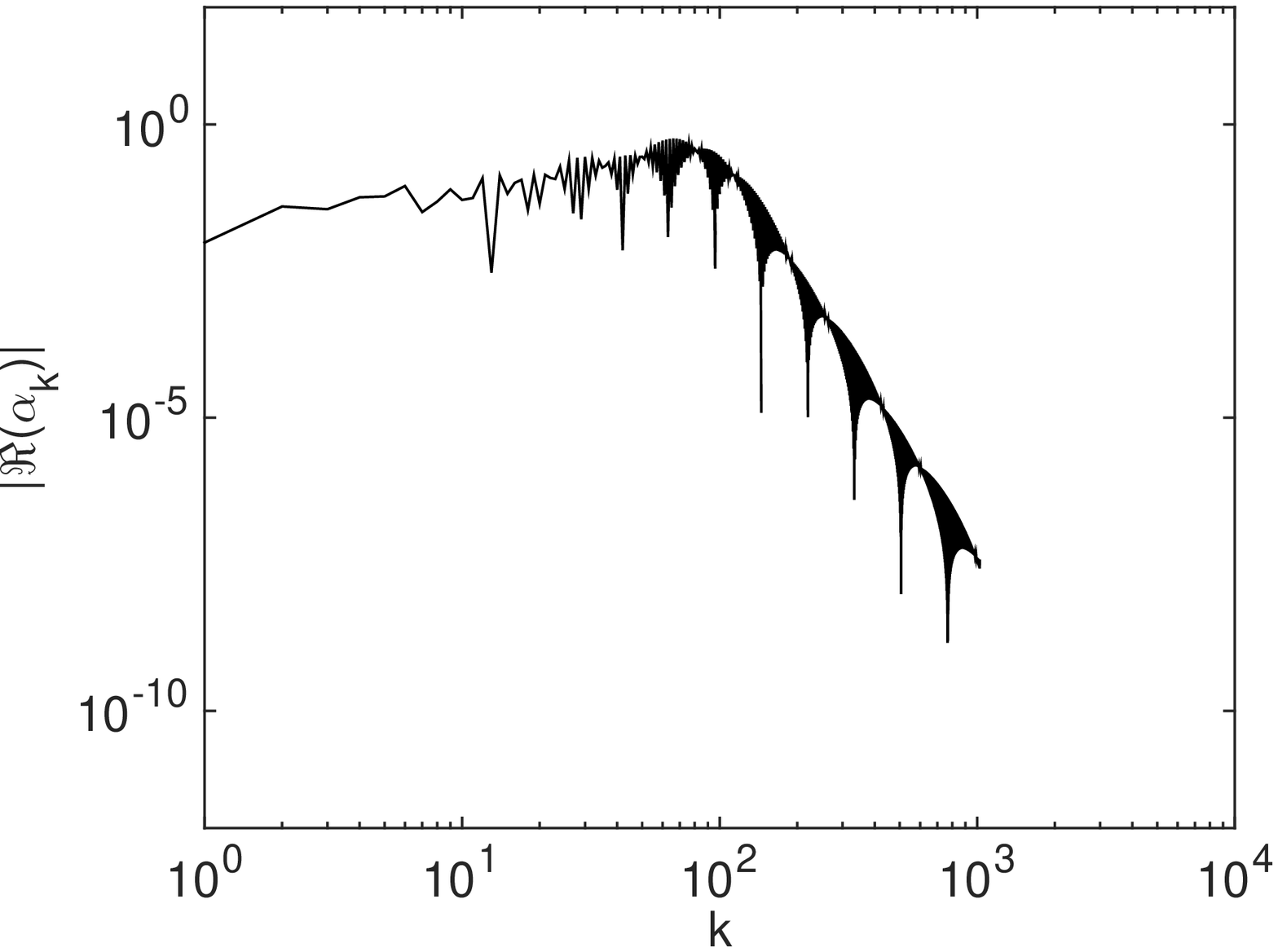}}\quad
\subfigure[$\delta = \frac{1}{128}$]
{\includegraphics[width=0.5\textwidth]{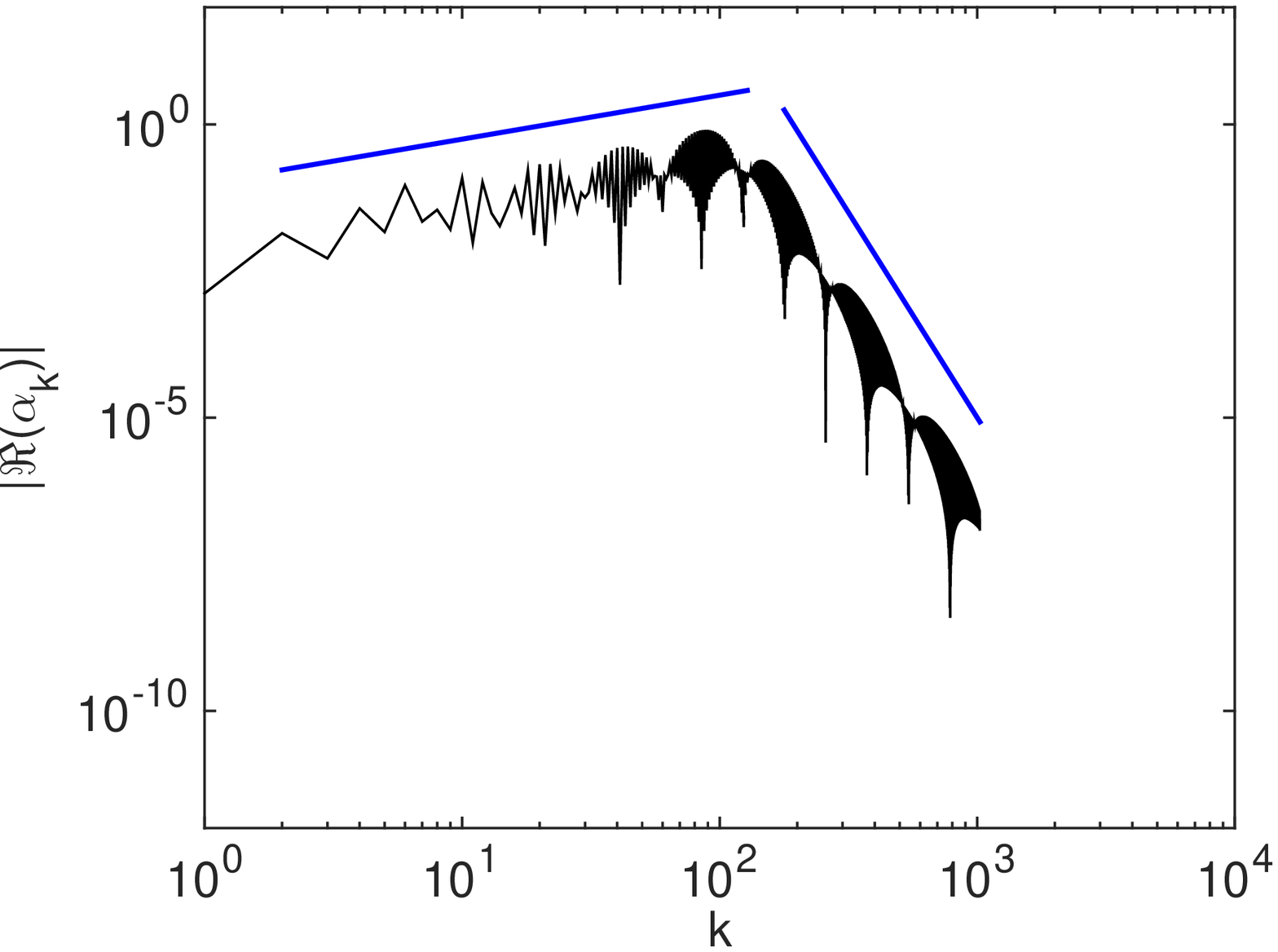}}
}
\end{center}
\caption{Fourier coefficient spectra ${|\Re(\alpha_k)|}$,
  $k=1,\dots,N$, of the neutrally-stable eigenvectors $u^N$ shown in
  figure \ref{fig:neut_th} for the indicated values of the
  regularization parameter $\delta$. In figure (d) the blue solid
  lines represent the slopes of $3/4$ and $-7$, respectively, for $k <
  1/\delta$ and for $k > 1 / \delta$.}
\label{fig:neut_sp}
\end{figure}

Finally, we go on to discuss the eigenvectors associated with the
purely real eigenvalues forming the continuous part of the spectrum.
Since, as demonstrated in figure \ref{fig:ReLam}, for increasing
resolutions $N$ different eigenvalues are actually computed in the
continuous spectrum, there is no sense of convergence with $N$. We
will therefore {analyze} here the effect of decreasing the
regularization parameter $\delta$ at a fixed resolution $N=1024$.
{As above, we will focus on the real parts of the eigenvectors
  (with the imaginary parts having similar properties).}  To
{fix} attention, we consider the neutrally-stable eigenvector
associated with the eigenvalue $\lambda \approx 1.0502$.  In figure
\ref{fig:neut_th} we show the dependence of ${\Re(u^N(\theta))}$
on the polar angle $\theta$ with $N=1024$ and for different values of
the regularization parameter $\delta$. We observe that as $\delta$
decreases the oscillations move away from the centre of the domain
$[0,\pi]$ towards the endpoints.  The number of oscillations, however,
remains approximately constant.  The corresponding Fourier coefficient
spectra are shown in figures \ref{fig:neut_sp}(a--d). We see in these
plots that the Fourier coefficients increase with $k$ as
$|\Re(\alpha_k)| \sim k^{3/4}$ when $k < 1 / \delta$ {which}
demonstrates that the neutrally-stable eigenvectors are actually more
singular objects than the stable and unstable eigenvectors discussed
above. We note that in the regularized regime (i.e., for $k > 1 /
\delta$) the Fourier coefficient spectrum of the neutrally-stable
eigenvectors vanishes with the slope of -7, see figure
\ref{fig:evec_kc_sp}(d).  Extrapolating from the trends evident in
figures \ref{fig:neut_th} and \ref{fig:neut_sp} to the limit $\delta
\rightarrow 0$, we can anticipate that the neutrally-stable
eigenfunctions will have the form of a finite number of oscillations
localized in an infinitesimal neighbourhood of the stagnation points
$\theta=0$ and $\theta=\pi$. The number of these oscillations appears
to be an increasing function of the eigenvalue magnitude $|\lambda|$.

\section{Discussion}
\label{sec:discuss}

{ In this section we first provide a simple argument to justify the
  numerical results obtained in \S \ref{sec:results} and then make
  some comparisons with the results of earlier studies. Some
  properties of the eigenvectors discussed in \S \ref{sec:results} are
  consequences of the ``degeneracy'' of the stability operator
  \eqref{eq:La}. More specifically, knowing the streamfunction field
  \eqref{eq:psiHill} characterizing Hill's vortex, the coefficient of
  the derivative term on the RHS in \eqref{eq:La} can be expressed as
  $\v_0\cdot\t_{\theta} = (\C a^2 / 5) \sin\theta$ which vanishes at
  the endpoints $\theta = 0, \pi$. {To illustrate the effect of
    this degeneracy we will consider a simplified model problem
    obtained from \eqref{eq:evalp1} by dropping the integral terms and
    rescaling the coefficients, so that we obtain
\begin{equation}
i\lambda\, w(\theta) = \sin\theta \frac{dw(\theta)}{d\theta}, \qquad
\theta \in [0,\pi]
\label{eq:sindudt}
\end{equation}
for some $w(\theta)$.  We now perform a change of variables $s=s(\theta)$
defined through $ds = d \theta / \sin\theta$, so that
\begin{equation}
s(\theta) = \int_{\pi/2}^\theta \frac{d\theta'}{\sin\theta'} =
\ln\left( \csc\theta - \cot\theta \right),
\label{eq:st}
\end{equation}
where the lower integration bound was chosen to make the
transformation antisymmetric with respect to the midpoint of the
interval $[0,\pi]$.  Transformation \eqref{eq:st} has the properties
$\lim_{\theta \rightarrow 0} s(\theta) = - \infty$ and $\lim_{\theta
  \rightarrow \pi} s(\theta) = + \infty$, such that it represents a
one-to-one map from the interval $[0,\pi]$ to the real line.
Introducing this change of variables in equation \eqref{eq:sindudt},
we obtain
\begin{equation}
i\lambda \, w(s) = \sin\theta \frac{dw(s)}{\sin\theta \, ds} = \frac{dw(s)}{ds}, 
\qquad s \in \RR
\label{eq:duds}.
\end{equation}
It then follows from \eqref{eq:st} and \eqref{eq:duds} that equation
\eqref{eq:sindudt} admits a continuous spectrum coinciding with the
entire complex plane $\lambda \in \CC$ with the eigenfunctions given
by $w(\theta) = e^{i\lambda s(\theta)}$. When the eigenvalues are
restricted to the real line $\lambda = \lambda_{\text{re}} \in \RR$,
the corresponding eigenfunctions $w(\theta)$ exhibit oscillations with
wavelengths decreasing as $\theta \rightarrow 0,\pi$, as was also
observed in \S \ref{sec:results} for the neutrally-stable modes,
cf.~figure \ref{fig:neut_th}. On the other hand, for purely imaginary
eigenvalues $\lambda = i \lambda_{\text{im}}$, where
$\lambda_{\text{im}} \in \RR$, the corresponding eigenfunctions take
the form $w(\theta) = \left( \csc\theta -
  \cot\theta\right)^{-\lambda_{\text{im}}}$ which for
$\lambda_{\text{im}} < 0$ has the properties $\lim_{\theta \rightarrow
  0^+} w(\theta) = \infty$ and $\lim_{\theta \rightarrow \pi^-}
w(\theta) = 0$ consistent with the singular behaviour of the unstable
eigenmodes observed in \S \ref{sec:results}, cf.~figures
\ref{fig:evec_N_th} and \ref{fig:evec_kc_th}. Thus, one can conclude
that the singular structure of the eigenvectors is a consequence of
the degeneracy of the coefficient in front of the derivative term in
\eqref{eq:La} and some qualitative insights about this issue can be
deduced based on the simplified problem \eqref{eq:sindudt}.} We add
that similar problems are known to arise in hydrodynamic stability,
for example, in the context of the inviscid Rayleigh equation
describing the stability of plane parallel flows
\citep{Schmid2001book}. In that problem, however, the singularity
appears inside the domain giving rise to critical layers with
locations dependent on the eigenvalues.}

{We now} return to a remark made in Introduction, namely, that
Hill's vortices represent a one-parameter family of solutions
parameterized by the constant $\C$, or equivalently, by the
translation velocity $W$, cf.~\eqref{eq:W}. {We remark that the
  stability operator defined in \eqref{eq:La} is linear in $\C$ which
  implies that eigenvalues $\lambda$ (and hence also the growth rates)
  will be proportional to $\C$ (or $W$). This is also} consistent with
the observations made by \citet{mm78}.

{Next we compare our findings with the results of \citet{mm78} which
  concerned essentially the same problem. We remark that these results
  were verified computationally by \citet{p86}.  The exponential
  growth rate of the unstable perturbations predicted by \citet{mm78}
  was (using our present notation) $-3W / a = 6 / 15 = 0.4$ which is
  in excellent agreement with the first unstable eigenvalue found
  here, cf.~figure \ref{fig:ImLam}(a). Similar agreement was found as
  regards the structure of the most unstable perturbation ---
  \citet{mm78} also found it to have the form of a localized spike at
  the rear stagnation point (the fact that this spike had a finite
  width seems related to the truncation of the infinite system of
  ordinary differential equations). It appears that the second
  unstable mode, cf.~figure \ref{fig:evec_N_th}(b), was undetected by
  the analysis of \citet{mm78} due to its smaller growth rate,
  cf.~figure \ref{fig:ImLam}(b).}

{To close this section, we comment on the continuous part of the
  spectrum which was reported in \S \ref{sec:results}, cf.~figures
  \ref{fig:sp} and \ref{fig:ReLam}. Such continuous spectra often
  appear in the study of infinite-dimensional Hamiltonian, or more
  generally, non self-adjoint systems, where nontrivial effects may
  arise from its interaction with the discrete spectrum \citep{m06}.
  In the present problem, however, the results of \citet{mm78} and
  \citet{p86} indicate that the observed instability has the form of a
  purely modal growth which can be completely explained in terms of
  the discrete spectrum and the associated eigenfunctions. Moreover,
  this is confirmed by the very good agreement between the growth rate
  of the instability determined by \citet{mm78} and the value of the
  first unstable eigenvalue obtained in our study. These observations
  thus allow us to conclude that there is no evidence for any role the
  continuous spectrum might play in the observed instability
  mechanism.}

\section{Conclusions}
\label{sec:final}

In this study we have considered the linear stability of Hill's vortex
with respect to axisymmetric {circulation-preserving}
perturbations. This was done using the systematic approach of
\citet{ep13} to obtain an eigenvalue problem characterizing the
linearized evolution of perturbations to the shape of the vortex
boundary. Recognizing that the Euler equation describing the evolution
of discontinuous vorticity distributions gives rise to a {\em
  free-boundary} problem, our approach was based on shape
differentiation of the contour-dynamics formulation in the 3D
axisymmetric geometry \citep{slf08}. As such, it did not involve the
simplifications invoked in the earlier studies of this problem by
\citet{mm78,frk94,r99} which were related to, e.g., approximate only
satisfaction of the kinematic conditions on the vortex boundary. The
resulting singular integro-differential operator was approximated with
a spectral method in which the integral expressions were evaluated
analytically and using {\tt chebfun}. We considered a sequence of
regularized eigenvalue problems \eqref{eq:Md} featuring smooth
eigenfunctions for which the convergence of the numerical
approximation was established. Then, the original problem was
recovered in the limit of vanishing regularization parameter $\delta$.
Since in the limit $\delta\rightarrow 0$ the eigenfunctions were found
to be distributions, the convergence of this approach with the
resolution $N$ was not very fast, but it did provide a precise
characterization of their regularity in terms of the rate of decay of
the Fourier coefficients in expansion \eqref{eq:un}.  Following this
procedure we showed that the stability operator has four purely
imaginary eigenvalues, associated with two unstable and two stable
eigenmodes, in addition to a continuous spectrum of purely real
eigenvalues associated with neutrally-stable eigenmodes. The two
unstable eigenmodes are distributions in the form of infinitely sharp
peaks localized at the rear stagnation point and differ by their
degree of singularity.  The stable eigenmodes have the form of similar
peaks localized at the front stagnation point. On the other hand, the
neutrally-stable eigenvectors have the form of ``wiggles''
concentrated in a vanishing neighbourhood of the two stagnation points
with the number of oscillations increasing with the eigenvalue
magnitude $|\lambda|$.

Our results are consistent with the findings from the earlier studies
of this problem by \citet{mm78,frk94,r99}. We emphasize that these
earlier studies did not, however, solve the complete linear stability
problem and only considered the linearized evolution of some
prescribed initial perturbation (they can be therefore regarded as
evaluating the action of an operator (matrix) on a vector, rather than
determining all of its eigenvalues and eigenvectors). These studies
did conclude that initial perturbations evolve towards a sharp peak
concentrated near the rear stagnation point. Thus, our present
findings may be interpreted as sharpening the results of these earlier
studies. {In particular, excellent agreement was found with the
  growth rate of the unstable perturbations found by \citet{mm78}.}

{The findings of the present study lead to some intriguing
  questions concerning the initial-value problem for the evolution of
  Hill's vortex with a perturbed boundary. It appears that, in the
  continuous setting without any regularization, this problem may not
  be well-posed, in the sense that, for generic initial perturbations,
  the vortex boundary may exhibit the same poor regularity as observed
  for the unstable eigenvectors in \S \ref{sec:results} (i.e., be at
  least discontinuous). While it is possible that the nonlinearity
  might exert some regularizing effect, this is an aspect of the
  problem which should be taken into account in its numerical solution.
  A standard numerical approach to the solution of such problems is
  the axisymmetric version of the ``contour dynamics'' method
  \citep{p86,wr96,slf08} in which the discretization of the contour
  boundary with straight segments or circular arcs combined with an
  approximation of the singular integrals provide the required
  regularizing effect. On the other hand, the singular structure of
  the solution can be captured more readily with higher-order methods,
  such as the spectral approach developed here.}

There is a number of related problems which deserve attention and will
be considered in the near future. A natural extension of the questions
addressed here is to investigate the stability of Hill's vortex with
respect to non-axisymmetric perturbations, as already explored by
\citet{frk94,r99}. Another interesting question is to consider the
effect of swirl \citep{moffatt69,hh10}. Hill's vortex is a member of
the Norbury-Fraenkel family of inviscid vortex rings and their
stability remains an open problem. It was argued by \citet{mm78} that
the highly localized nature of the boundary response of Hill's vortex
to perturbations is a consequence of the presence of a stagnation
point.  Since the Norbury-Fraenkel vortices other than Hill's vortex
do not feature stagnation points {on the vortex boundary}, it may
be conjectured that in those cases eigenfunctions of the stability
operator will be smooth functions of the arclength coordinate.
{Therefore, in the context of the linear stability problem, the
  family of the Norbury-Fraenkel vortex rings may be regarded as a
  ``regularization'' of Hill's vortex analogous and alternative to our
  approach developed in \S \ref{sec:numer},
  cf.~\eqref{eq:F}--\eqref{eq:Md}. The different problems mentioned in
  this paragraph, except for the effect of swirl,} can be investigated
using the approach developed by \citet{ep13} and also employed in the
present study. {As regards the stability of Hill's vortex with
  swirl, the difficulty stems from the fact that, to the best of our
  knowledge, there is currently no vortex-dynamics formulation of the
  type \eqref{eq:v} available for axisymmetric flows with swirl. Our
  next step will be to analyze the stability of the Norbury-Fraenkel
  vortex rings to axisymmetric perturbations.}  Finally, it will
{also} be interesting to compare the present findings with the
results of the short-wavelength stability analysis of \citet{hh10}. In
particular, one would like to know if there is any overlap between the
two stability analyses and, if so, whether they can produce comparable
predictions of the growth rates.

\section*{Acknowledgments}

The authors are thankful to Toby Driscoll for helpful advice on a
number of {\tt chebfun}-related issues {and to Dmitry Pelinovsky
  for his comments on the singular structure of the eigenfunctions.
  Anonymous referees provided constructive comments which helped us to
  improve the paper.} B.P. acknowledges the support through an NSERC
(Canada) Discovery Grant.



\begin{thebibliography}{47}
\expandafter\ifx\csname natexlab\endcsname\relax\def\natexlab#1{#1}\fi

\bibitem[Baker(1990)]{b90}
{\sc Baker, G.~R.} 1990 A study of the numerical stability of the method of
  contour dynamics. {\em Phil. Trans. Roy. Soc.\/} {\bf 333}, 391--400.

\bibitem[Bliss(1973)]{b73}
{\sc Bliss, D.~B.} 1973 The dynamics of flows with high concentrations of
  vorticity. PhD thesis, Department of Aeronautics and Astronautics,
  Massachusetts Institute of Technology.

\bibitem[Delfour \& Zol\'esio(2001)]{dz01a}
{\sc Delfour, M.~C. \& Zol\'esio, J.-P.} 2001 {\em Shape and Geometries ---
  Analysis, Differential Calculus and Optimization\/}. SIAM.

\bibitem[Drazin \& Reid(2004)]{Drazin2004book}
{\sc Drazin, P.G. \& Reid, W.H.} 2004 {\em Hydrodynamic stability\/}, 2nd edn.
  Cambridge, UK: Cambridge University Press.

\bibitem[Driscoll {\em et~al.\/}(2014)Driscoll, Hale \& Trefethen]{chebfun}
{\sc Driscoll, T.~A., Hale, N. \& Trefethen, L.~N.} 2014 {\em Chebfun Guide\/},
  {Pafnuty Publications} edn. Oxford.

\bibitem[Dritschel(1985)]{d85}
{\sc Dritschel, D.~G.} 1985 The stability and energetics of corotating uniform
  vortices. {\em J. Fluid Mech.\/} {\bf 157}, 95--134.

\bibitem[Dritschel(1988)]{d88}
{\sc Dritschel, D.~G.} 1988 Nonlinear stability bounds for inviscid,
  two-dimensional, parallel or circular flows with monotonic vorticity, and the
  analogous three-dimensional quasi-geostrophic flows. {\em J. Fluid Mech.\/}
  {\bf 191}, 575--581.

\bibitem[Dritschel(1990)]{d90}
{\sc Dritschel, D.~G.} 1990 The stability of elliptical vortices in an external
  straining flow. {\em J. Fluid Mech.\/} {\bf 210}, 223--261.

\bibitem[Dritschel(1995)]{d95}
{\sc Dritschel, D.~G.} 1995 A general theory for two--dimensional vortex
  interactions. {\em J. Fluid Mech.\/} {\bf 293}, 269--303.

\bibitem[Dritschel \& Legras(1991)]{dl91}
{\sc Dritschel, D.~G. \& Legras, B.} 1991 {The elliptical models of
  two-dimensional vortex dynamics. II: Disturbance equations}. {\em Phys.
  Fluids A\/} {\bf 3}, 855--869.

\bibitem[Elcrat {\em et~al.\/}(2005)Elcrat, Fornberg \& Miller]{efm05}
{\sc Elcrat, A., Fornberg, B. \& Miller, K.} 2005 Stability of vortices in
  equilibrium with a cylinder. {\em J. Fluid Mech.\/} {\bf 544}, 53--68.

\bibitem[Elcrat \& Protas(2013)]{ep13}
{\sc Elcrat, A. \& Protas, B.} 2013 A framework for linear stability analysis
  of finite-area vortices. {\em Proceedings of the Royal Society A\/} {\bf
  469}, 20120709.

\bibitem[Fraenkel(1972)]{fraenkel-1972-JFM}
{\sc Fraenkel, L.~E.} 1972 Examples of steady vortex rings of small
  cross--section in an ideal fluid. {\em J. Fluid Mech.\/} {\bf 51}, 119.

\bibitem[Fukumoto \& Moffatt(2008)]{fm08}
{\sc Fukumoto, Y. \& Moffatt, H.~K.} 2008 Kinematic variational principle for
  motion of vortex rings. {\em Physica D\/} {\bf 237}, 2210--2217.

\bibitem[Fukuyu {\em et~al.\/}(1994)Fukuyu, Ruzi \& Kanai]{frk94}
{\sc Fukuyu, A., Ruzi, T. \& Kanai, A.} 1994 The response of {Hill's} vortex to
  a small three dimensional disturbance. {\em J. Phys. Soc. Japan\/} {\bf 63},
  510--527.

\bibitem[Guo {\em et~al.\/}(2004)Guo, Hallstrom \& Spirn]{ghs04}
{\sc Guo, Y., Hallstrom, Ch. \& Spirn, D.} 2004 {Dynamics Near an Unstable
  Kirchhoff Ellipse}. {\em Commun. Math. Phys.\/} {\bf 245}, 297--354.

\bibitem[Hackbusch(1995)]{h95}
{\sc Hackbusch, W.} 1995 {\em Integral Equations: Theory and Numerical
  Treatment\/}. Birkh\"auser.

\bibitem[Hattori \& Hijiya(2010)]{hh10}
{\sc Hattori, Y. \& Hijiya, K.} 2010 Short-wavelength stability analysis of
  {Hill's} vortex with/without swirl. {\em Phys Fluids\/} {\bf 22}, 074104.

\bibitem[Hill(1894)]{hill-1894}
{\sc Hill, M. J.~M.} 1894 On a spherical vortex. {\em Philos. Trans. Roy. Soc.
  London\/} {\bf A185}, 213--245.

\bibitem[Kamm(1987)]{k87}
{\sc Kamm, J.~R.} 1987 Shape and stability of two--dimensional vortex regions.
  PhD thesis, Caltech.

\bibitem[Kelvin(1880)]{k80}
{\sc Kelvin, Lord} 1880 Vibrations of a columnar vortex. {\em Phil. Mag.\/}
  {\bf 10}, 155--168.

\bibitem[Laub(2005)]{l05}
{\sc Laub, A.~J.} 2005 {\em Matrix Analysis for Scientists and Engineers\/}.
  SIAM.

\bibitem[Lifschitz(1995)]{l95}
{\sc Lifschitz, A.} 1995 nstabilities of ideal fluids and related topics. {\em
  Z. Angew. Math. Mech.\/} {\bf 75}, 411.

\bibitem[Lifschitz \& Hameiri(1991)]{lh91}
{\sc Lifschitz, A. \& Hameiri, E.} 1991 Local stability conditions in fluid
  dynamics. {\em Phys. Fluids A\/} {\bf 3}, 2644--2651.

\bibitem[{Llewellyn Smith} \& Ford(2001)]{lsf01}
{\sc {Llewellyn Smith}, S.~G. \& Ford, R.} 2001 {Three-dimensional acoustic
  scattering by vortical flows. Part I: General theory}. {\em Phys. Fluids\/}
  {\bf 13}, 2876--2889.

\bibitem[Love(1893)]{l93}
{\sc Love, A. E.~H.} 1893 On the stability of certain vortex motions. {\em
  Proc. London Math. Soc.\/} {\bf s1--25}, 18--43.

\bibitem[Luzzatto-Fegiz \& Williamson(2010)]{fw10a}
{\sc Luzzatto-Fegiz, P. \& Williamson, C. H.~K.} 2010 Stability of conservative
  flows and new steady-fluid solutions from bifurcation diagrams exploiting
  variational argument. {\em Phys. Rev. Lett.\/} {\bf 104}, 044504.

\bibitem[Luzzatto-Fegiz \& Williamson(2012)]{fw12b}
{\sc Luzzatto-Fegiz, P. \& Williamson, C. H.~K.} 2012 Determining the stability
  of steady two-dimensional flows through imperfect velocity-impulse diagrams.
  {\em J. Fluid Mech.\/} {\bf 706}, 323--350.

\bibitem[Mitchell \& Rossi(2008)]{mr08}
{\sc Mitchell, T.~B. \& Rossi, L.~F.} 2008 The evolution of {Kirchhoff}
  elliptic vortices. {\em Phys. Fluids\/} {\bf 20}, 054103.

\bibitem[Moffatt(1969)]{moffatt69}
{\sc Moffatt, H.~K.} 1969 The degree of knottedness of tangled vortex lines.
  {\em J. Fluid Mech.\/} {\bf 35}, 117--129.

\bibitem[Moffatt \& Moore(1978)]{mm78}
{\sc Moffatt, H.~K. \& Moore, D.~W.} 1978 The response of {Hill's} spherical
  vortex to a small axisymmetric disturbance. {\em J. Fluid Mech.\/} {\bf 87},
  749--760.

\bibitem[Mohseni(2001)]{mohseni-2001-PF}
{\sc Mohseni, K.} 2001 Statistical equilibrium theory for axisymmetric flow:
  Kelvin's variationalprinciple and an explanation for the vortex ring
  pinch--off process. {\em Phys. Fluids\/} {\bf 13}, 1924.

\bibitem[Moore \& Saffman(1971)]{ms71}
{\sc Moore, D.~W. \& Saffman, P.~G.} 1971 Structure of a line vortex in an
  imposed strain. In {\em Aircraft wake turbulence\/} (ed. Goldburg Olsen \&
  Rogers), pp. 339--354. Plenum.

\bibitem[Norbury(1973)]{norbury-1973-JFM}
{\sc Norbury, J.} 1973 A family of steady vortex rings. {\em J. Fluid Mech.\/}
  {\bf 57}, 417--431.

\bibitem[Olver {\em et~al.\/}(2010)Olver, Lozier, Boisvert \& Clark]{olbc10}
{\sc Olver, F.~W.~J., Lozier, D.~W., Boisvert, R.~F. \& Clark, C.~W.}, ed. 2010
  {\em {NIST Handbook of Mathematical Functions}\/}. New York, NY: Cambridge
  University Press.

\bibitem[Pozrikidis(1986)]{p86}
{\sc Pozrikidis, C.} 1986 The nonlinear instability of {Hill's} vortex. {\em J.
  Fluid Mech.\/} {\bf 168}, 337 -- 367.

\bibitem[Pullin(1992)]{p92}
{\sc Pullin, D.~I.} 1992 Contour dynamics methods. {\em Annual Review of Fluid
  Mechanics\/} {\bf 24}, 89--115.

\bibitem[Rozi(1999)]{r99}
{\sc Rozi, T.} 1999 {Evolution of the Surface of Hill's Vortex Subjected to a
  Small Three-Dimensional Disturbance for the Cases of $m=0,2,3$ and 4}. {\em
  J. Phys. Soc. Japan\/} {\bf 68}, 2940.

\bibitem[Rozi \& Fukumoto(2000)]{rf00}
{\sc Rozi, T. \& Fukumoto, Y.} 2000 {The Most Unstable Perturbation of
  Wave-Packet Form Inside Hill's Vortex}. {\em J. Phys. Soc. Japan\/} {\bf 69},
  2700--2701.

\bibitem[Saffman(1992)]{saffman-1992}
{\sc Saffman, P.~G.} 1992 {\em Vortex Dynamics\/}. Cambridge, New York:
  Cambridge University Press.

\bibitem[Saffman \& Szeto(1980)]{ss80}
{\sc Saffman, P.~G. \& Szeto, R.} 1980 Equilibrium shapes of a pair of equal
  uniform vortices. {\em Phys Fluids\/} {\bf 23}, 2339--2342.

\bibitem[Schmid \& Hennigson(2001)]{Schmid2001book}
{\sc Schmid, P.~J.\ \& Hennigson, D.~S.} 2001 {\em Stability and Transition in
  Shear Flows\/}. New York: Springer-Verlag.

\bibitem[Shariff {\em et~al.\/}(2008)Shariff, Leonard \& Ferziger]{slf08}
{\sc Shariff, K., Leonard, A. \& Ferziger, J.~H.} 2008 A contour dynamics
  algorithm for axisymmetric flow. {\em Journal of Computational Physics\/}
  {\bf 227}, 9044--9062.

\bibitem[Wakelin \& Riley(1996)]{wr96}
{\sc Wakelin, S.~L. \& Riley, N.} 1996 Vortex ring interactions ii. inviscid
  models. {\em Quarterly Journal of Mechanics and Applied Mathematics\/} {\bf
  49}, 287--309.

\bibitem[Weinstein(2006)]{m06}
{\sc Weinstein, M.I.} 2006 {Extended Hamiltonian Systems}. In {\em Handbook Of
  Dynamical Systems\/} (ed. B.~Hasselblatt \& A.~Katok), , vol.~1B.
  North-Holland.

\bibitem[Wu {\em et~al.\/}(1984)Wu, II \& Zabusky]{woz84}
{\sc Wu, H.~M., II, E. A.~Overman \& Zabusky, N.~J.} 1984 {Steady-State
  Solutions of the Euler Equations in Two Dimensions: Rotating and Translating
  V-States and Limiting Cases. I. Numerical Algorithms and Results}. {\em
  J.~Comp.~Phys\/} {\bf 53}, 42--71.

\bibitem[Wu {\em et~al.\/}(2006)Wu, Ma \& Zhou]{wu-book-2006}
{\sc Wu, J.-Z., Ma, H.-Y. \& Zhou, M.-D.} 2006 {\em Vorticity and vortex
  dynamics\/}. Berlin, Heidelberg, New York: Springer-Verlag.

\end{thebibliography}

\end{document}